\documentclass[twocolumn]{aastex631}

\usepackage[T1]{fontenc} 
\usepackage[utf8]{inputenc} 

\usepackage{inputenc}
\usepackage{amssymb}
\usepackage{graphicx}
 \usepackage{ulem}

\accepted{September 21, 2024}

\begin{document}


\title{\textit{Searching for GEMS:} TOI-6383Ab, a giant planet transiting an M3-dwarf star in a binary system\footnote{Based on observations obtained with the Hobby-Eberly Telescope (HET), which is a joint project of the University of Texas at Austin, the Pennsylvania State University, Ludwig-Maximillians-Universitaet Muenchen, and Georg-August Universitaet Goettingen. The HET is named in honor of its principal benefactors, William P. Hobby and Robert E. Eberly.}}

\author[0000-0002-8035-1032]{Lia Marta Bernab\`o}
\affiliation{Institute of Planetary Research, German Aerospace Center (DLR), Rutherfordstrasse 2, 12489 Berlin}
\affiliation{McDonald Observatory and Department of Astronomy, The University of Texas at Austin, USA}
\author[0000-0001-8401-4300]{Shubham Kanodia} \affiliation{Earth and Planets Laboratory, Carnegie Science, 5241 Broad Branch Road, NW, Washington, DC 20015, USA}
\author[0000-0003-4835-0619]{Caleb I. Ca$\mathrm{\tilde{N}}$as}\affiliation{NASA Goddard Space Flight Center, 8800 Greenbelt Road, Greenbelt, MD 20771, USA}
\author[0000-0001-9662-3496]{William D. Cochran}\affiliation{McDonald Observatory and Department of Astronomy, The University of Texas at Austin, USA}\affiliation{Center for Planetary Systems Habitability, The University of Texas at Austin, USA}
\author[0000-0001-6803-9698]{Szil\'ard Csizmadia}\affiliation{Institute of Planetary Research, German Aerospace Center (DLR), Rutherfordstrasse 2, 12489 Berlin}\affiliation{ELKH-SZTE Stellar Astrophysics Research Group, H-6500 Baja, Szegedi \'ut Kt. 766, Hungary}
\author[0000-0001-9596-7983]{Suvrath Mahadevan}\affiliation{Department of Astronomy \& Astrophysics, 525 Davey Laboratory, The Pennsylvania State University, University Park, PA 16802, USA}\affiliation{Center for Exoplanets and Habitable Worlds, 525 Davey Laboratory, The Pennsylvania State University, University Park, PA 16802, USA}\affiliation{ETH Zurich, Institute for Particle Physics \& Astrophysics, Switzerland}
\author[0000-0001-7409-5688]{Gudhmundur Stef{\'A}nsson}\affiliation{Anton Pannekoek Institute for Astronomy, University of Amsterdam, Science Park 904, 1098 XH Amsterdam, The Netherlands}
\author[0000-0002-5463-9980]{Arvind F. Gupta}\affiliation{U.S. National Science Foundation National Optical-Infrared Astronomy Research Laboratory, 950 N. Cherry Ave., Tucson, AZ 85719, USA}
\author[0000-0002-0048-2586]{Andrew Monson}\affiliation{Steward Observatory, The University of Arizona, 933 N. Cherry Avenue, Tucson, AZ 85721, USA}
\author[0000-0002-4475-4176]{Henry A. Kobulnicky}\affiliation{Department of Physics \& Astronomy, University of Wyoming, Laramie, WY 82070, USA}
\author[0000-0002-2401-8411]{Alexander K. Larsen}\affiliation{Department of Physics \& Astronomy, University of Wyoming, Laramie, WY 82070, USA}
\author[0009-0003-1637-8315]{Ethan G. Cotter}\affiliation{Department of Physics \& Astronomy, University of Wyoming, Laramie, WY 82070, USA}
\author{Alexina Birkholz}\affiliation{Department of Physics \& Astronomy, University of Wyoming, Laramie, WY 82070, USA}
\author[0000-0002-5817-202X]{Tera N. Swaby}\affiliation{Department of Physics \& Astronomy, University of Wyoming, Laramie, WY 82070, USA}
\author[0000-0003-2307-0629]{Gregory Zeimann}\affiliation{Hobby Eberly Telescope, University of Texas at Austin, TX 78712, USA}
\author[0000-0003-4384-7220]{Chad F. Bender}\affiliation{Steward Observatory, The University of Arizona, 933 N. Cherry Avenue, Tucson, AZ 85721, USA}
\author[0000-0002-2144-0764]{Scott A. Diddams}\affiliation{Electrical, Computer \& Energy Engineering, University of Colorado, 425 UCB, Boulder, CO 80309, USA}\affiliation{Department of Physics, University of Colorado, 2000 Colorado Avenue, Boulder, CO 80309, USA}
\author[0000-0002-2990-7613]{Jessica E. Libby-Roberts}\affiliation{Department of Astronomy \& Astrophysics, 525 Davey Laboratory, The Pennsylvania State University, University Park, PA 16802, USA}\affiliation{Center for Exoplanets and Habitable Worlds, 525 Davey Laboratory, The Pennsylvania State University, University Park, PA 16802, USA}
\author[0000-0002-9082-6337]{Andrea S.J. Lin}\affiliation{Department of Astronomy \& Astrophysics, 525 Davey Laboratory, The Pennsylvania State University, University Park, PA 16802, USA}\affiliation{Center for Exoplanets and Habitable Worlds, 525 Davey Laboratory, The Pennsylvania State University, University Park, PA 16802, USA}
\author[0000-0001-8720-5612]{Joe P. Ninan}\affiliation{Department of Astronomy and Astrophysics, Tata Institute of Fundamental Research, Homi Bhabha Road, Colaba, Mumbai 400005, India}
\author[0000-0002-6510-1828]{Heike Rauer}\affiliation{Institute of Planetary Research, German Aerospace Center (DLR), Rutherfordstrasse 2, 12489 Berlin}\affiliation{Institut f{\"u}r Geologische Wissenschaften, Freie Universit{\"a}t Berlin, 12249 Berlin, Germany}
\author[0009-0006-7298-619X]{Varghese Reji}\affiliation{Department of Astronomy and Astrophysics, Tata Institute of Fundamental Research, Homi Bhabha Road, Colaba, Mumbai 400005, India}
\author[0000-0003-0149-9678]{Paul Robertson}\affiliation{Department of Physics \& Astronomy, University of California Irvine, Irvine, CA 92697, USA}
\author[0000-0001-8127-5775]{Arpita Roy}\affiliation{Astrophysics \& Space Institute, Schmidt Sciences, New York, NY 10011, USA}
\author[0000-0002-4046-987X]{Christian Schwab}\affiliation{School of Mathematical and Physical Sciences, Macquarie University, Balaclava Road, North Ryde, NSW 2109, Australia}

\correspondingauthor{Lia Marta Bernabò}
\email{lia.bernabo@dlr.de \\ liamarta.bernabo@gmail.com}



\begin{abstract}

We report on the discovery of a transiting giant planet around the 3500 K M3-dwarf star TOI-6383A located 172 pc from Earth. It was detected by the Transiting Exoplanet Survey Satellite (TESS) and confirmed by a combination of ground-based follow-up photometry and precise radial velocity measurements. This planet has an orbital period of $\sim$1.791 days, mass of 1.040$\pm$0.094 $M_J$ and a radius of 1.008$^{+0.036}_{-0.033} ~R_J$, resulting in a mean bulk density of 1.26$^{+0.18}_{-0.17}$ g cm$^{-3}$. TOI-6383A has an M-dwarf companion star, TOI-6383B, which has a stellar effective temperature $T_{eff}$ $\sim$ 3100 K and a projected orbital separation of 3100 AU. TOI-6383A is a low-mass dwarf star hosting a giant planet and is an intriguing object for planetary evolution studies due to its high planet-to-star mass ratio. This discovery is part of the \textit{Searching for Giant Exoplanets around M-dwarf Stars (GEMS)} Survey, intending to provide robust and accurate estimates of the occurrence of GEMS and the statistics on their physical and orbital parameters. This paper presents an interesting addition to the small number of confirmed GEMS, particularly notable since its formation necessitates massive, dust-rich protoplanetary discs and high accretion efficiency ($>$ 10\%).
\end{abstract}



\section{Introduction} \label{sec:intro}
The formation of giant planets is expected to follow one of two main pathways: core accretion or gravitational instability. The core accretion scenario (\citealp{Mizuno1980, Pollack1996}) is a bottom-up process. Small solid particles coagulate in the protoplanetary disc, followed by the gradual growth of the planetary embryo. A phase of rapid gas accretion follows, allowing the core to acquire a massive gaseous envelope. The runaway gas accretion phase can only be triggered with a massive core ($\gtrsim$ 10 M$_\Earth$). This is hypothesized to take place in the Class II stage of planet formation, though simulations (\citealp{Laughlin2004, Burn2021}) suggested that this mechanism for forming giant planets might suffer from a problem of mass-budget and formation timescales around M-dwarfs. Theoretical models predict that the timescale for the formation of Jupiter-like gas giants could exceed the lifespan of the gas disc (see \citealp{Laughlin2004} and references therein), hinting at the possibility of alternative processes in the formation of massive gas giant planets around M-dwarf stars. \\
\indent On the other hand, gravitational instability theory of planet formation (\citealp{Boss1997}) consists of a rapid formation mechanism that involves the break-up of a massive protostellar disc into clumps under their self-gravity. These clumps subsequently undergo contraction and collapse to give rise to giant planets or substellar objects. In the outer regions of the disc, far beyond the ice line, gravitational instability has been hypothesized to be able to form giant planets around M-dwarfs (\citealp{Boss2006, Boss2011, BossKanodia2023,Mercer2020}). The formation is then supposed to be followed by migration because of the interaction with the disc. This scenario takes place during the Class 0 or I disc phase, quite early in the life of the disc when it is much more massive (\citealp{Tychoniec2020}).

\noindent Characterizing and cataloguing the growing sample of Giant Exoplanets around M-dwarf Stars (GEMS; transiting with planetary radius $\gtrsim$8 R$_\Earth$, orbiting M-dwarf stars with T$_\mathrm{eff} \leq$ 4000 K) may reconcile the existence of this sample with the two contemporary theories of giant planet formation. This is the goal of the {\it Searching for GEMS survey} (\citealp{Kanodia2024}) aiming to explore recent discoveries of GEMS and assess them in light of existing theories of planet formation. \cite{Kanodia2024} carried out an exhaustive discussion on the occurrence and formation of GEMS and showed that approximately 40 GEMS with 5$\sigma$ mass measurements are required to better quantify the significance of potential trends between stellar mass and bulk density for giant planets. The required sample size is double the current $\sim$20 planet sample. Identified candidates are undergoing concurrent validation and follow-up observations by ground-based telescopes (photometric and spectroscopic surveys) to confirm their planetary status and determine their stellar and planetary characteristics.

\noindent In this manuscript, we present the detection of a giant planet around the M3 spectral type dwarf star TOI-6383A. The discovery was made using TESS, and a ground-based program of observations followed to characterize the system. In particular, we obtained six ground-based transits with Red Buttes Observatory (RBO; \citealp{Kasper2016}), in Wyoming, USA, along with precise Radial Velocities (RVs) from the Habitable-zone Planet Finder (HPF; \citealp{Mahadevan2012,Mahadevan2014}). TOI-6383A has companion M-dwarf at a projected distance of $\sim$18 $^{\prime \prime}$ (3100 AU) away. TOI-6383B has a M5 spectral type and an effective temperature of 3100 K.\\
\indent The paper is structured as follows: in Section \ref{sec:observations} we describe the photometric and spectroscopic observations of TOI-6383A; in Section \ref{sec:stellar_parameters} we report our investigations on the characteristics of the host star and its companion star TOI-6383B; in Section \ref{sec:analysis}, we describe the joint transit and RV fitting procedure and the results on the planetary and orbital parameters and in Section \ref{sec:discussion} we put the planet in context and discuss its interior and formation.

\section{Observations}
\label{sec:observations}

\subsection{Transit light curves}
\label{subsec:LCs}
\subsubsection{TESS}
\begin{figure}
    \centering
\includegraphics[width=0.48\textwidth]{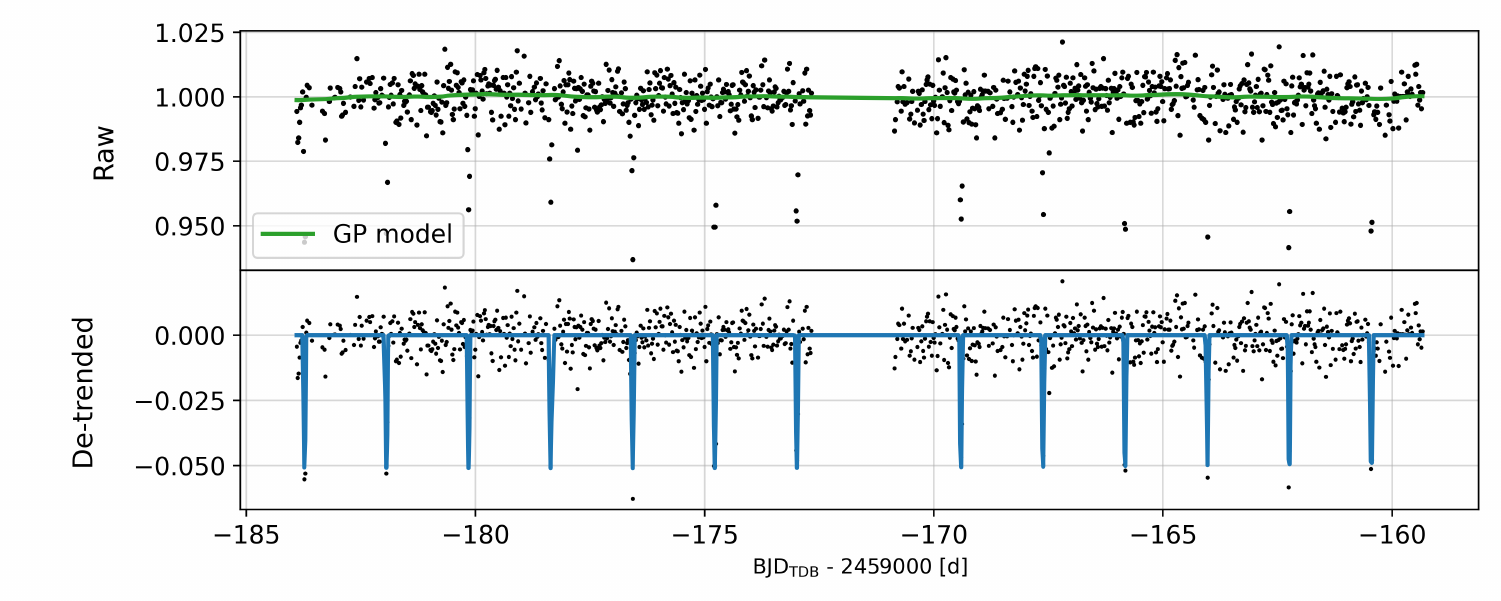}
\includegraphics[width=0.48\textwidth]{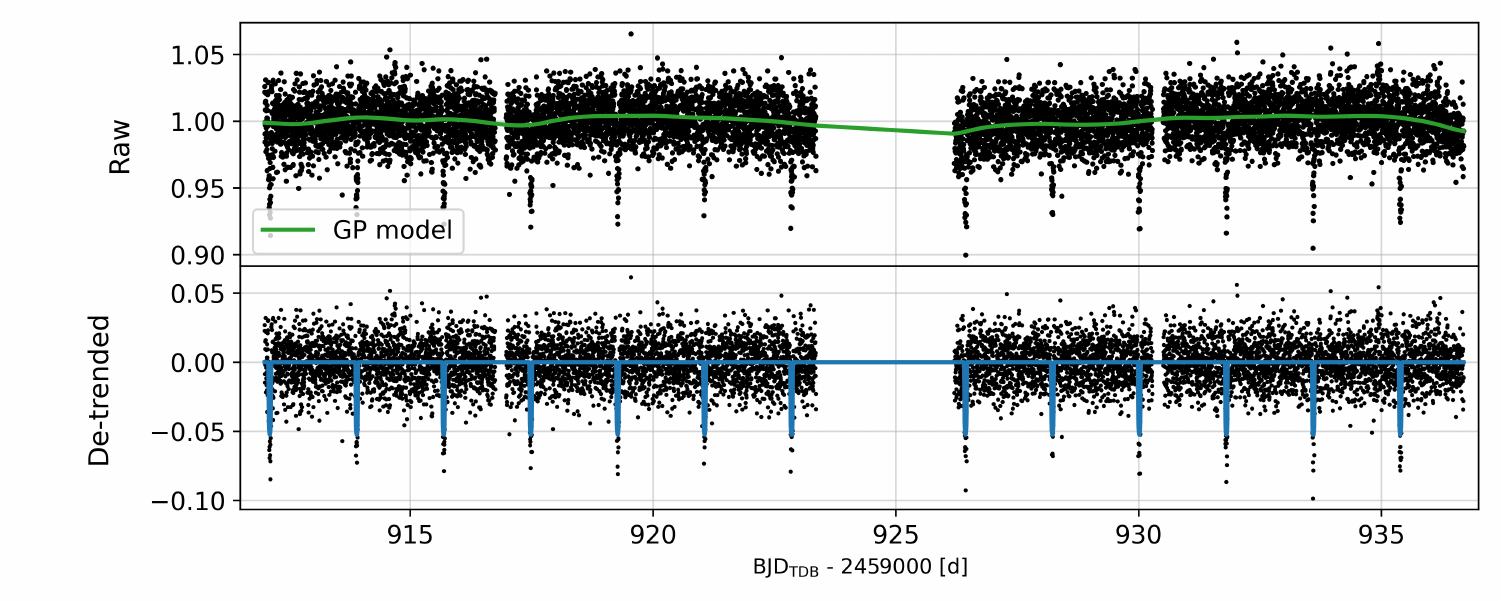}
    \caption{TESS transits observed in Sectors 19 and 59 excluding masked datapoints, modelled with Gaussian Processes (GP). The exposure times are 1800 s and 200 s, respectively.}
    \label{fig:TESS_LC_sectors}
\end{figure}

TOI-6383A (TIC-328513434, {\it Gaia} Data Release 3 473934011733049856, RA $\alpha_{J2016}$= 04:01:41.93, Dec $\delta_{J2016}$= +04:03:33.93) was observed by TESS (\citealp{TESS2015}) in Sector 19 in Camera 2 starting in the night of 28 December 2019 and observing the target for 27 days. The images were acquired with an 1800 s exposure time. Starting on 26 November 2022 TOI-6383A was re-observed in Sector 59 in Camera 2 with a 200 s exposure time. We generate the light curve (LC) from the TESS full-frame images (FFIs) by employing the open-source tool \texttt{eleanor} (\citealp{Feinstein2019}). This software utilizes the \texttt{TESScut} service to capture a 31 × 31 pixel excerpt from the calibrated FFIs, centered on TOI-6383A. The light curve is computed from the \texttt{CORR\_FLUX} values, employing linear regression with factors such as pixel position, background measurement, and time to eliminate correlated signals. We use the default \texttt{eleanor} aperture, resulting in a differential photometric precision (CDPP) of 4332 ppm and 10014 ppm for Sectors 19 and 59, respectively. \\ 
\indent The \texttt{eleanor} aperture includes numerous field stars, contributing to the photometric dilution observed in the TESS light curve. This necessitates subsequent ground-based follow-up to resolve these background stars. \\
\indent Figure \ref{fig:TESS_LC_sectors} shows the raw and de-trended TESS photometry of both Sectors, modelled with Gaussian Process (GP) regression, which is described more in Section \ref{sec:analysis}.\\
\indent 

\noindent We perform ground-based photometric and spectroscopic follow-up to confirm the presence of the planet and characterize its properties. 

\subsubsection{Ground-based photometric observations}
Between August and November 2023, we observed six transits of TOI-6383Ab with the 0.6 m telescope located at the Red Buttes Observatory (RBO; \citealp{Kasper2016}), Wyoming, USA. The telescope is an f/8.43 Ritchey–Chrétien Cassegrain, currently equipped with an Apogee Alta F16 camera, and constructed by DFM Engineering, Inc. The transits were observed with the Bessell I filter (\citealp{Bessel1990}) and 240 s exposure times and a field of view of 8.94 × 8.94$^{\prime}$.
Table \ref{tab:photometry_RBO} shows a summary of the main properties of the data acquisition of the photometric follow-up program. The FWHM of the source in RBO photometry ranges from approximately 1.8 to 3.2 pixels, while the scale is 0.72 $^{\prime \prime}$/pixel. Therefore aperture sizes of RBO photometry are enough to estimate and correct for dilution in the TESS photometry after careful aperture selection and avoid contamination of the 18$^{\prime\prime}$ away companion star TOI-6383B in the RBO photometry.\\
\indent The RBO data are processed using the \texttt{Python} package \texttt{astropy} (\citealp{Astropy2013}) to perform bias, dark and flat field processing. Astrometry information is verified (or added if missing) using the \texttt{astrometry.net} package (\citealp{Hogg2008}). Aperture photometry is performed using \texttt{Python photutils} (\citealp{Bradley2023}) at the locations (corrected for proper motions) of all {\it Gaia} sources (brighter than G = 17 mag using \texttt{astroquery}) with multiple apertures at once. Using the {\it Gaia} coordinates allows for easy and consistent identification and cross-matching. The midpoints of the exposures are converted to BJD$_\mathrm{TDB}$ using \texttt{barycorrpy} (\citealp{KanodiaWright2018}), a \texttt{Python} implementation of algorithms by \cite{Eastman2012}. To perform relative photometry, the flux from the target star is compared to the sum of the fluxes of the selected comparison stars which are chosen to: be present on all frames, have no nearby companions, and not be known variables. The final choice of aperture was made by minimizing the scatter in the data. Further refinement of comparison stars is done by manually (de)selecting stars to yield consistent light curves across all defined apertures. These light curves are shown after modelling in Figure \ref{fig:LCs_fitted} along with the two TESS transits. 
\begin{table}[]
    \centering
    \begin{tabular}{c c c c}
    \hline \hline
     Obs. date & PSF FWHM & Airmass\\
     UT (DD-MM-YY) & min - max [$^{\prime \prime}$] & start - mid - end \\ 
     \hline
     21-08-2023 & 1.84 - 2.54 & 2.02 - 1.34 - 1.10\\
     30-08-2023 & 1.88 - 2.75 & 2.15 - 1.41 - 1.16\\
     08-09-2023 & 2.08 - 3.20 & 2.30 - 1.33 - 1.08 \\
     26-09-2023 & 1.79 - 2.87 & 2.66 - 1.59 - 1.17 \\
     30-10-2023 & 1.80 - 2.96 & 1.55 - 1.09 - 1.16 \\
     15-11-2023 & 1.83 - 2.32 & 1.13 - 1.07 - 1.23 \\
     \hline
     \hspace{0.3cm}
    \end{tabular}
    \caption{Summary of the main properties of the ground-based follow-up program at RBO. In the fourth column, we report the minimum and maximum values of the FWHM during the night. In the last column, we report the airmass at the beginning, in the middle of the observations and at the end of the night.}
    \label{tab:photometry_RBO}
\end{table}

\subsection{RV follow-up with HPF}
\label{subsec:RVs}
We also acquired 10 radial velocity visits of TOI-6383A with the Habitable-zone Planet Finder (HPF; \citealp{Mahadevan2012, Mahadevan2014,Mahadevan2018}), a near-infrared, stabilized (\citealp{Stefansson2016}), fiber-fed (\citealp{Kanodia2018}), high-resolution (R$\sim$50~000) radial velocity spectrograph (\citealp{Metcalf2019}) located at the 10-meter Hobby-Eberly Telescope (HET; \citealp{Ramsey1998,Hill2021}) at McDonald Observatory in Texas, USA. 
The instrument bandpass covers the range from 8080 to 12780~\AA. \\
\indent The high-resolution spectra were acquired between August and October 2023. Ten visits were made to the target to obtain a full-phase RV curve. Each visit consists of two exposures of 969 s each, where the Signal-to-Noise (S/N) per pixel per unbinned exposure ranges between a minimum of $\sim$21 and a maximum of $\sim$32 at 1070 nm. The two RV datapoints acquired each night are binned by weighted averaging to improve the S/N. The errorbar is then calculated as the harmonic mean of the squared errorbars on the single measurements. The binned RVs are reported in Table~\ref{tab:RVs} and shown phase-folded in Figure~\ref{fig:RV_phasefolded}.\\
\indent We use the algorithms described in the tool {\tt HxRGproc} (\cite{Ninan2018}) for bias removal, non-linearity correction, cosmic ray correction, slope/flux and variance image calculation of the raw HPF data.\\
\indent We utilize the template-matching method (e.g., \citealp{AngladaEscudeButler2012}) to derive radial velocities from the collected spectra. This method is incorporated within the \texttt{SpEctrum Radial Velocity AnaLyser} pipeline (\texttt{SERVAL}; \citealp{Zechmeister2018}), customized for HPF (\citealp{Metcalf2019}). Initially, a master template is constructed based on all HPF observations of TOI-6383A. Subsequently, for each individual observation, we ascertain the Doppler shift by adjusting its velocity to minimize the $\chi^2$ when compared against the template. The telluric regions are identified using a synthetic telluric-line mask generated by \texttt{telfit} (\citealp{Gullikson2014}), a Python interface for the Line-by-Line Radiative Transfer Model package (\citealp{Clough2005}). After excluding the telluric and sky-emission lines, the master template is formed using HPF observations of the target star. To compensate for the barycentric motion in the individual spectra, we use the barycentric correction algorithms of \cite{WrightEastman2014}, implemented in the \texttt{Python} package \texttt{barycorrpy} (\citealp{KanodiaWright2018}).

\begin{table}[]
    \centering
    \begin{tabular}{c c c}
    \hline \hline
     Time & RV & $\sigma_\mathrm{RV}$\\
     BJD$_\mathrm{TDB}$ - 2~460~000 [d] & [m/s] & [m/s]\\
     \hline
182.954575 & -352 & 32  \\ 
187.938320 & -38 & 43 \\ %
190.910754 & 165 & 45 \\ %
192.921425 & 180 & 31 \\ %
194.900860 & 52 & 53 \\ %
196.893194 & -157 & 60 \\ %
197.894106 & 58 & 47 \\ %
215.868385 & 112 & 46 \\ %
215.995449 & 167 & 40 \\ %
225.986554 & -408 & 41 \\ %
     \hline
     \hspace{0.3cm}
    \end{tabular}
    \caption{Binned radial velocity datapoints of TOI-6383A acquired by HPF: time in BJD$_\mathrm{TDB}$, RV and errorbar on RV.}
    \label{tab:RVs}
\end{table}

\subsection{High-contrast Imaging: NESSI}

\begin{figure}
    \centering
    \includegraphics[width=0.48\textwidth]{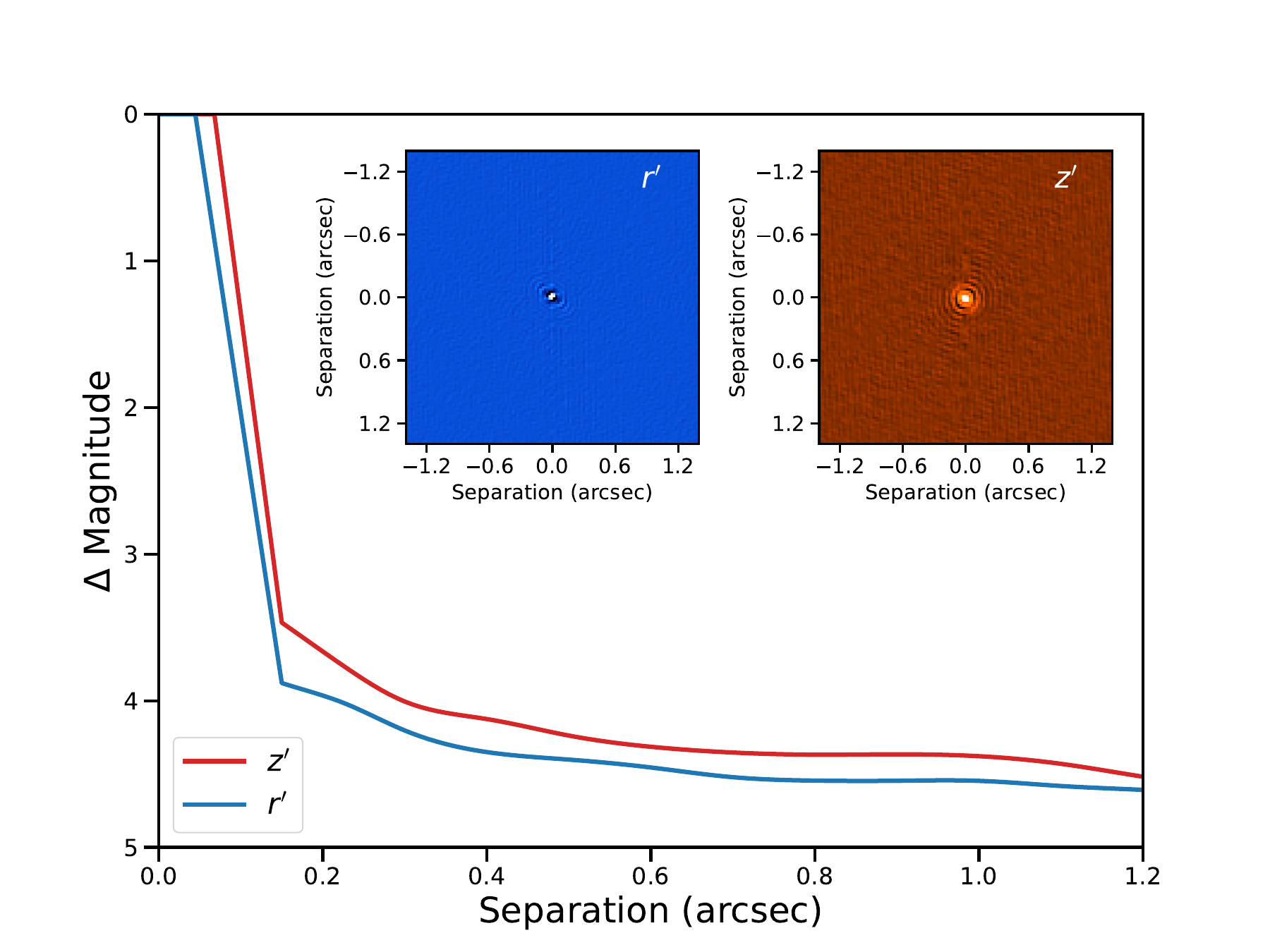}
    \caption{Reconstructed NESSI speckle images and 5$\sigma$ contrast curves for TOI-6383A. Observations were taken simultaneously in the {\it r'} filter with the blue camera (upper left inset image) and the {\it z'} filter with the red camera (upper right inset image). The contrast curves indicate the limiting magnitude difference at which bound or background sources could be detected for separations between 0.2$^{\prime \prime}$ and 1.2$^{\prime \prime}$. Note that the stellar companion falls outside the field of view.}
    \label{fig:NESSI}
\end{figure}

TOI-6383A was observed with the NN-Explore Exoplanet Stellar Speckle Imager (NESSI; \citealp{Scott2018}) on the WIYN Observatory\footnote{The WIYN Observatory is a joint facility of the NSF's National Optical-Infrared Astronomy Research Laboratory, Indiana University, the University of Wisconsin-Madison, Pennsylvania State University, Purdue University and Princeton University.}  on 9 September 2023. Sequences of diffraction-limited frames were taken in the SDSS {\it r'} and {\it z'} filters and processed following the methods described in \cite{Howell2011} to create high-resolution, reconstructed speckle images. We show the reconstructed images along with the achieved contrast limits as a function of angular separation in Figure \ref{fig:NESSI}. The NESSI results exclude the presence of nearby sources at $5\sigma$ confidence down to relative magnitude limits of $\Delta r' = 3.96$ mag and $\Delta z' = 3.66$ mag at a separation of 0.2$^{\prime \prime}$ and limits of $\Delta r' = 4.55$ mag and $\Delta z' = 4.38$ mag at a separation of 1.2$^{\prime \prime}$.

\begin{table*}[]
\small
\resizebox{\textwidth}{!}{
\centering
\begin{tabular}{l l c c c}
\hline \hline
    Parameter & Symbol [units] & TOI-6383A & TOI-6383B & Reference \\
    \hline
{\bf Identifiers}\\
TESS Input Catalogue & TIC & 328513434 & 328513444 & \\
TESS Object of Interest & TOI & 6383.01 & & \\
2MASS & & J04014183+6053297 & J04014299+6053457& \\  
{\it Gaia} identifier & DR2 & 473934011733049856 & 473934007432818688 & GAIA DR2\\
WISE & & J040141.89+605329.2 & J040143.05+605345.3 & WISE$^{\dagger 1}$\\
\\
{\bf Coordinates}\\
Right Ascension & RA ($\alpha_{J2016}$) [h:m:s] & 04:01:41.9293 $\pm$ 00:00:00.0015 & 04:01:43.0906 $\pm$ 00:00:00.0082 & GAIA DR3 \\ 
Declination & Dec ($\delta_{J2016}$) [h:m:s] & 
+04:03:33.9267 $\pm$ 00:00:00.0014 & +04:03:34.9980 $\pm$ 00:00:00.0070 & GAIA DR3 \\
Apparent separation & [\arcsec] & ... & 18.16552$\pm$ 0.00023 & GAIA DR3\\
Parallax & $\varpi$ [mas] & $5.847\pm0.031$ & $6.10\pm0.16$ & GAIA DR3\\
Distance from Earth & d [pc] & $172.08^{+0.91}_{-0.82}$ & $182.09^{+5.5}_{-6.2}$ & GAIA DR3\\
Projected physical separation & [AU] & ... & $3126\pm17$ & GAIA DR3 \\
Proper motion & pmRA [mas yr$^{-1}$] & $41.023\pm0.035$ & $41.02\pm0.17$ & GAIA DR3\\
Proper motion & pmDec [mas yr$^{-1}$] & $-51.30\pm0.032$ & $-51.28\pm0.16$ & GAIA DR3\\
\\
{\bf Magnitudes}\\
Johnson B & B [mag] & 18.52 $\pm$ 0.16 & ... & APASS$^{\dagger 2}$\\
Johnson V & V [mag] & 16.63 $\pm$ 0.20 & ... & APASS\\
\textit{J} & \textit{J} [mag] & 12.985 $\pm$ 0.026 & 15.174 $\pm$ 0.053 & 2MASS \\ 
H & [mag] & 12.318 $\pm$ 0.017 & 14.775 $\pm$ 0.069 & 2MASS \\
$K_s$ & [mag] & 12.087 $\pm$ 0.021 & 14.222 $\pm$ 0.071 & 2MASS \\
{\it Gaia} & G [mag] & 15.66280 $\pm$ 0.00067 & 18.5239 $\pm$ 0.0018 & GAIA DR3 \\
TESS & \textit{T} [mag] & 14.4506 $\pm$ 0.0073 & 17.064 $\pm$ 0.010 & TESS \\
WISE 3.4 $\mu$m & W1 [mag] & 11.967 $\pm$ 0.024 & 14.134 $\pm$ 0.029 & WISE \\
WISE 4.6 $\mu$m & W2 [mag] & 11.860 $\pm$ 0.022 & 13.945 $\pm$ 0.038& WISE \\
WISE 12 $\mu$m & W3 [mag] & 11.61 $\pm$ 0.20 & 12.715 & WISE \\
WISE 22 $\mu$m & W4 [mag] & 9.184 & 8.678 & WISE \\
\\
{\it Gaia} $B_P-R_P$ colour & G$_\mathrm{BP}$-G$_\mathrm{RP}$ & 2.5110 $\pm$ 0.0089 & 3.310 $\pm$ 0.076  & GAIA DR3 \\
\\
{\bf Stellar parameters}\\
Radius$^a$ & R$_\star$ [R$_\Sun$] & 0.457 $\pm$ 0.019 & 0.217$^{+0.017}_{-0.019}$ & This work \\ 
Mass$^b$ & M$_\star$ [M$_\Sun$] & 0.458 $\pm$ 0.011 & 0.205 $\pm$ 0.008 & This work \\
Mean density & $\rho_\star$ [g cm$^{-3}$] & 6.74 $\pm$ 0.35 & 25.62 $\pm$ 5.59 & This work \\
Effective temperature & T$_\mathrm{eff}$ [K] & 3444 $\pm$ 88 $^c$ & 3121 $\pm$ 81 $^d$ & This work \\
Surface gravity$^c$ & log$_{10}$(g) [cgs] & 4.81 $\pm$ 0.05 & ... & This work \\
Luminosity & L$_\star$ [L$_\Sun$] 
& 0.0326 $\pm$ 0.0009 & 0.0042 $\pm$ 0.0005 & This work \\
Stellar type & & M3$^{e, f}$ & M5$^e$ & This work \\
\\

{\bf Other stellar parameters}\\
V sin (i)$^c$ & [km s$^{-1}$] & <2.0  & ... & This work \\
Absolute radial velocity & [km s$^{-1}$] & 11.9 $\pm$ 0.3 & ... & GAIA DR3 \\
Galactic velocities (barycentric r. s.) & $U$, $V$, $W$ [km s$^{-1}$] & -40.07 $\pm$ 1.11 , -35.57 $\pm$ 1.44, -7.53 $\pm$ 0.31 & ... & This work \\ 
\hspace{0.5cm} " \hspace{0.5cm} " \hspace{0.78cm} (LSR) & $U$, $V$, $W$ [km s$^{-1}$] & -28.97 $\pm$ 1.39 , -21.33 $\pm$ 1.59 , -0.28 $\pm$ 0.68 & ...  & This work \\
\hline
\hspace{0.3cm}
\end{tabular}
    }
\caption{Stellar parameters for both TOI-6383A and TOI-6383B. All reported magnitudes are apparent.\\
Three dots "..." indicate that the parameter is not calculated or not available. \\
$^{\dagger 1}$ Wide-field Infrared Survey Explorer, \cite{WISE}\\
$^{\dagger 2}$ AAVSO Photometric All-Sky Survey, \cite{APASS_DR10}\\
$^a$ From its relation with the absolute $K_s$ magnitude, \cite{Mann2015}, see Section \ref{subsec:stellar_parameters};\\
$^b$ Equation (6) from \cite{Schweitzer2019}, based on the empirically calibrated sample, see Section \ref{subsec:stellar_parameters}; \\
$^c$ Calculated with \texttt{HPF-SpecMatch} algorithm, \cite{Stefansson2020}, see Section \ref{subsec:stellar_parameters};\\
$^d$ Equation (11) from \cite{Rabus2019}; \\
$^e$ From {\it Gaia} colours relations from \cite{Kiman2019}, as described in Section \ref{sec:spectral_classification}; \\
$^f$ Estimated with LRS-2 spectra, see Section \ref{sec:spectral_classification}. \\
}
    \label{tab:stellar_parameters}
\end{table*}

\subsection{LRS-2}
\label{sec:LRS-2}
To verify the spectral type and stellar characteristics of TOI-6383A, we conducted observations of the target using the second generation Low Resolution Spectrograph (LRS-2; \citealp{Lee2010, Chonis2016}) at the Hobby-Eberly Telescope located at McDonald Observatory in Texas, USA. LRS-2 is an optical integral field unit spectrograph with low resolution (R $\sim$1900), comprising two arms that simultaneously capture two fields of view measuring 6$^{\prime \prime}$ × 12$^{\prime \prime}$ each, with a separation of 100$^{\prime \prime}$. The red arm (LRS-2-R) comprises two channels covering $\sim$6430–8450~\AA~and $\sim$8230--10560~\AA, while the blue arm (LRS-2-B) is equipped with a pair of channels covering spectral ranges of $\sim$3640–4670~\AA~and $\sim$4640-7000~\AA. We collected spectra with LRS-2-B on 17 February 2024, with a seeing of 1.4$^{\prime \prime}$ and exposure times of 1800 s.\\
\indent The raw LRS-2 data are initially processed with \texttt{Panacea}\footnote{\url{https://github.com/grzeimann/Panacea}}, which carries out bias
subtraction, dark subtraction, fiber tracing, fiber wavelength evaluation, fiber extraction, fiber-to-fiber normalization, source detection, source extraction, and flux calibration for each channel. The absolute flux calibration comes from default response curves and measures of the mirror illumination
as well as the exposure throughput from guide images.\\
\indent Following the initial reduction process, we employed \texttt{LRS2Multi}\footnote{\url{https://github.com/grzeimann/LRS2Multi}} for advanced reduction steps and calibration of \texttt{Panacea} products. We pinpointed the target star, defined a 3.5$^{\prime \prime}$ aperture, and utilized fibers beyond this aperture to construct our sky model for each exposure. Subsequently, we subtracted the initial sky and generated a principal component basis consisting of 25 components from the residuals to eliminate further sky residuals caused by variable spectral Point Spread Functions (PSFs) for each fiber. The target spectrum was then extracted from the sky-subtracted frames 
and the resulting LRS-2 spectra of TOI-6383A were utilized to estimate the star's spectral type, as described in Section \ref{sec:spectral_classification}.

\section{Stellar parameters}
\label{sec:stellar_parameters}

\subsection{A companion star: TOI-6383B}

We searched the catalogue of the second and third {\it Gaia} data release (DR2, DR3; \citealp{GAIA,GAIADR2,GAIADR3} for possible co-moving companions to TOI-6383A. We searched all objects within 60$^{\prime\prime}$ of the target, which corresponds to a physical separation of around 10$^4$ AU at the distance of TOI-6383A. We selected the only object with comparable proper motions (PM), {\it Gaia} DR2 ID 473934007432818688. The star is $\sim$3 magnitudes fainter in the $G$-band, has parallax and coordinates $\mathrm{PM_{\alpha}}$ and $\mathrm{PM_{\delta}}$ which only slightly differ from that of TOI-6383A. They are listed in Table \ref{tab:stellar_parameters}. This putative M-dwarf companion is found at a separation of $\sim$18$^{\prime\prime}$ from TOI-6383A, which corresponds to a spatial projected separation of $\sim$ 3100 AU. \\
\indent TOI-6383A and TOI-6383B are present also in the catalogue of \cite{ElBadry2018} for {\it Gaia} Data Release 2, implying that the line-of-sight difference between the two objects in distance is less than twice their projected separation (within 3$\sigma$ limit). Moreover, they have proper motion differences within 3$\sigma$ of the maximum velocity difference expected for a system of total mass 5 M$_\Sun$ with circular orbits, meaning that the proper motion of the two stars is consistent with a bound Keplerian orbit.\\
\indent Based on the parallax and absolute $K_s$ magnitude, TOI-6383B is lower in mass than our planet host TOI-6383A, and hence referred to as TOI-6383B from now on.


\subsection{TOI-6383A and B stellar parameters}
\label{subsec:stellar_parameters}

The stellar parameters are derived using a combination of {\it Gaia} astrometry, LRS-2 spectra and photometric relations, and are summarized in Table \ref{tab:stellar_parameters}.\\
\indent The separation of the two stars in the binary system is far enough that TOI-6383A is cleanly resolved in the 2MASS (Two Micron All Sky Survey; \citealp{2MASS}) images, we could use the photometric relations to derive the mass and radius of both TOI-6383A and TOI-6383B. In more detail, we use the absolute $K_s$ magnitude to derive the radii (\citealp{Mann2015}), and from them, the stellar masses using the empirically calibrated sample from \cite{Schweitzer2019}. As indicated in \cite{Mann2015}, the errorbars on the radii are incremented by 2.89\% to account for the systematic scatter. These parameters are reported in Table \ref{tab:stellar_parameters}.\\ 

\indent We use \texttt{HPF-SpecMatch}\footnote{\url{https://gummiks.github.io/hpfspecmatch/}} routine (\citealp{Stefansson2020}) to calculate the stellar parameters from HPF
spectra. The template-matching method is based on \texttt{SpecMatch-Emp} from \cite{Yee2017} and compares the target spectrum with the HPF spectral library, containing 166 stars in the following parameter ranges for effective temperature T$_\mathrm{eff}$=[2700 K, 6000 K], surface gravity log (g) = [4.3, 5.3] and metallicity [Fe/H]=[-0.5, +0.5] dex. The spectrum of the target is compared with each spectrum from the library, returning its $\chi^2$. This determines the library stars ranking, as a first comparison method. Afterwards, only the top five library stars with the best $\chi^2$ values are taken into consideration. The $\chi^2$ metric is then employed to allocate scaling constants to each of these five best-fit library stars. This process results in the creation of a composite spectrum that closely aligns with the target spectrum. The scaling constants are instrumental in determining a weighted average, which facilitates precise parameter estimations for the target star's spectroscopic parameters. Following \cite{Stefansson2020}, we use a leave-one-out cross-validation approach to determine the errorbars on these parameters. In this process, a library star of interest is removed from the rest of the stellar library pool. Then, the algorithm is executed again to infer its stellar parameters independently of its true parameters. The differences between these calculated parameters and the true parameter values are used to estimate the uncertainty of inferred parameters. \\
\indent The calculated stellar parameters for TOI-6383A from HPF spectra are reported in Table \ref{tab:stellar_parameters}. Since we obtained no separate HPF observation of TOI-6383B, its effective temperature is calculated from its relation with the {\it Gaia} G magnitude (Equation (11) from \citealp{Rabus2019}).\\
\indent Due to the HPF spectral resolution limit, we can only determine an upper limit on {\it v} sin {\it i} for TOI-6383A. We also determine metallicity using two methodologies but note that the determination of the metallicity of M-dwarfs is challenging and has many caveats. Due to the lower temperature of the atmospheres of M-dwarf stars, they are characterized by molecular features. Unfortunately, this introduces complexities in determining their line profiles and metallicity.\\
\indent The first method makes use of with \texttt{HPF-SpecMatch}, which resulted in [Fe/H]= 0.15 $\pm$ 0.12 dex. However, the estimate is not reliable because of the multiplicity of the minima in the $\chi^2$ minimisation, making the parameter space "flat" with the stellar metallicity. \\
\indent We also estimated the metallicity of TOI-6383A using {\it Gaia} colours and the online tool \texttt{METaMorPHosis}\footnote{\url{https://chrduque.shinyapps.io/metamorphosis/}} \textemdash~provided by \cite{DuqueArribas2023} \textemdash~which derives several photometric calibrations of metallicity. We used $G$ absolute {\it Gaia} magnitude, $B_P-R_P$ {\it Gaia} colour and $J$, $H$ and $K$ 2MASS absolute magnitudes, as reported in Table \ref{tab:stellar_parameters}. The weighted mean among three estimates made with $B_P-R_P$-G, $B_P-R_P$-J and $B_P-R_P$-H planes is [Fe/H] = 0.20 $\pm$ 0.08 dex, which is consistent with the \texttt{HPF-SpecMatch} estimate within the errorbars. \\

\noindent We computed the Lomb-Scargle periodogram on the stellar photometry to check the presence of stellar rotation. We applied it on the {\it Zwicky Transient Facility} (ZTF; \citealp{Bellm2019, Graham2019}), with 521 photometric measurements with the {\it r} filter spanning $\sim$ 2000 days and
on the {\it All-Sky Automated Survey for Supernovae} (ASAS-SN; \citealp{Shappee2014,ASASSN2019}) survey 164 available {\it V} magnitude measurements spanning $\sim$ 1400 days. We could not find any periodicity above the 10\% False Alarm Probability limit.\\

\noindent In the context of galactic kinematics, we calculated the $U$, $V$ and $W$ velocities of TOI-6383A in the barycentric and Local (LSR) standard of rest frames using the \texttt{python} package \texttt{galpy} (\citealp{galpy}) and the relations from \cite{Schoenrich2010}. They are reported in Table \ref{tab:stellar_parameters}. Using the criterion from \cite{Bensby2014}, we can classify TOI-6383A as a thin disc star.

\begin{figure*}[]
    \centering
    \includegraphics[width=0.9\textwidth]{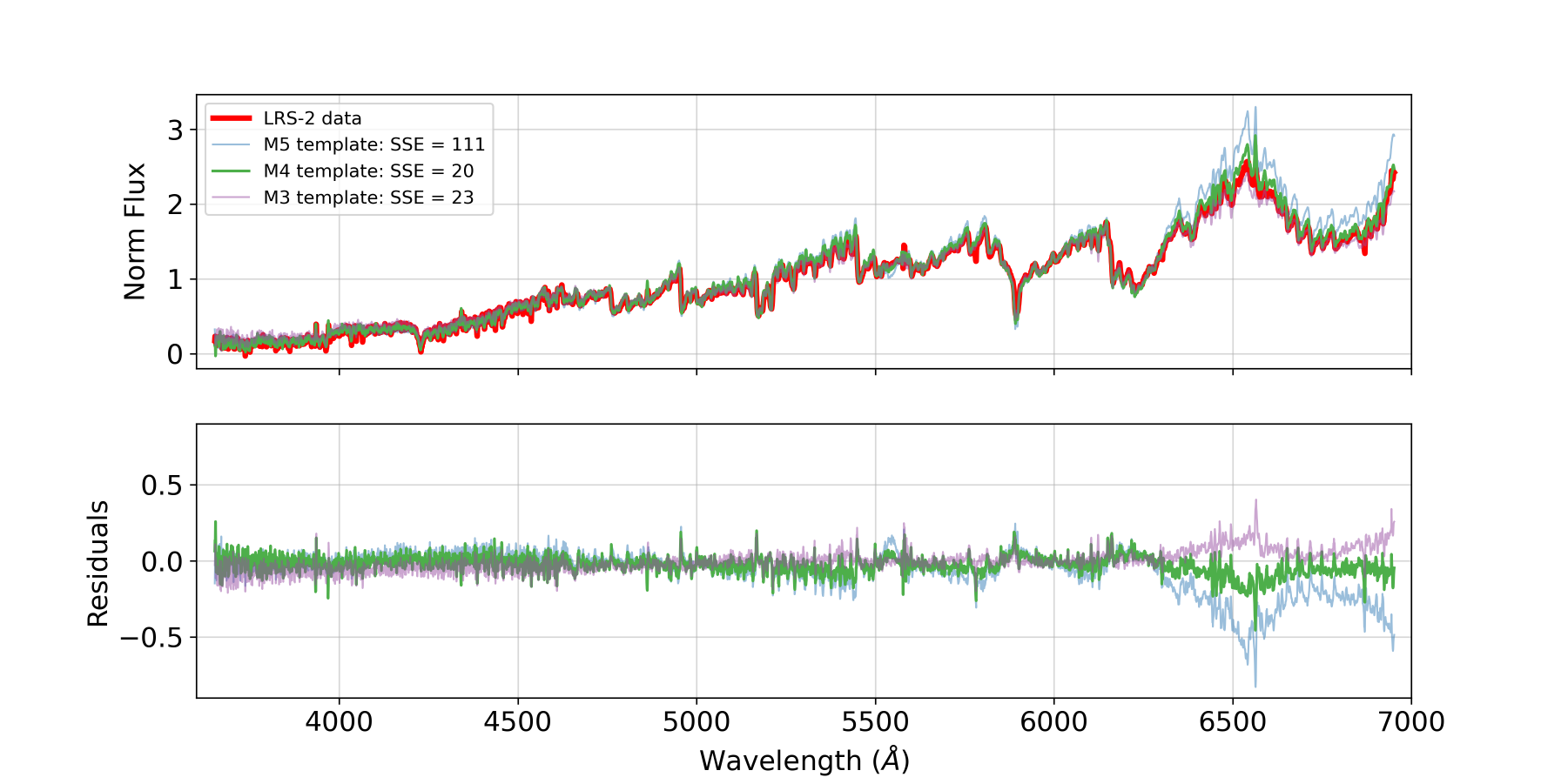}
    \caption{Comparison of the LRS-2 spectra with the empirical templates from \texttt{pyHammer}. The observed LRS-2-B spectra after response and telluric correction are plotted in red. Empirical templates of M3, M4 and M5 stars are also shown. The summed square errors (SSE) are reported in the legend. Along with the residuals (bottom part of the Figure), they show that the M3-4 stellar spectra are the best-matching ones.}
    \label{fig:LRS2}
\end{figure*}

\subsection{Spectral classification}
\label{sec:spectral_classification}

\indent To compute the spectral classification of TOI-6383A with the LRS-2 spectra, we used the \texttt{Python} package \texttt{pyHammer} (\citealp{Kesseli2017,Roulston2020}), which is based on \texttt{The Hammer} (\citealp{Covey2007}) and estimates of spectral type, metallicity and radial velocity, and is also to visually classify stellar spectra. We compare the observed spectrum with spectral line index measurements and template optical spectral templates. These are derived from the MaNGA Stellar Library (MaStar), consisting of calibrated optical spectra from Sloan Digital Sky Survey IV (\citealp{Yan2019}). From this comparison, we obtained a spectral type M3-4 (see Figure \ref{fig:LRS2}, and, in particular, the comparison between the residuals in the $\sim$6200--7000~\AA ~range in the bottom plot).\\

\noindent We also compare {\it Gaia} colours $B_P-R_P$, $B_P-G$ and $G-R_P$ with Table 4 of \cite{Kiman2019} to obtain the spectral type of both stars, obtaining M3 spectral type for TOI-6383A and M5-6 for TOI-6383B.\\
\indent We therefore adopt M3 as the spectral type of TOI-6383A, given that it matches the results from \cite{Kiman2019} and that the summed square errors of M3 and M4 templates are similar.

\section{Data analysis: joint fit of LCs and RVs}
\label{sec:analysis}
\begin{figure*}
    \centering
    \includegraphics[width=\textwidth]{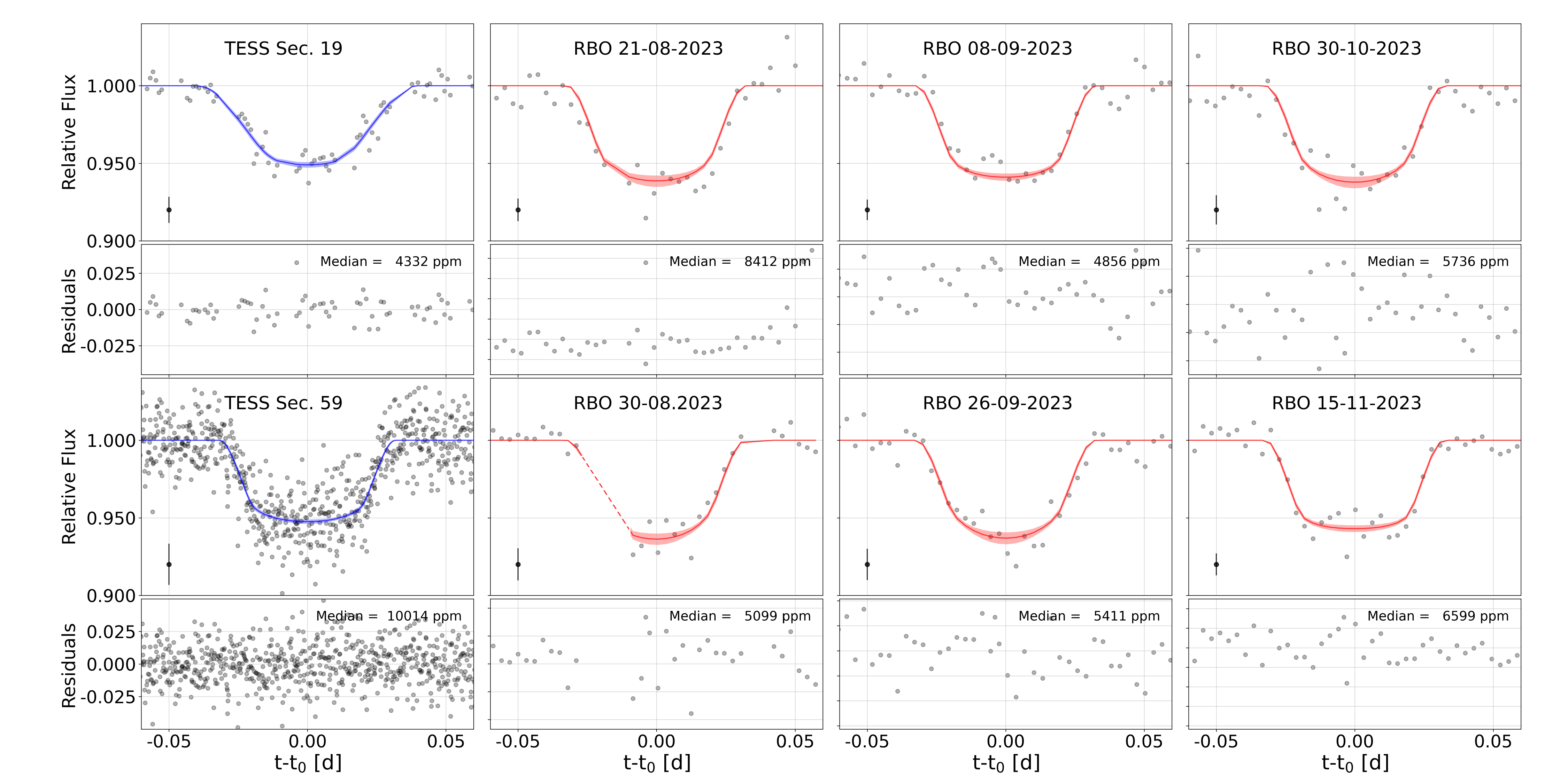}
    \caption{TESS and RBO light curves fitted by \texttt{exoplanet} jointly with RVs, with a focus on the transit region. Time is subtracted from the transit mid-time for each light curve. In all plots, the detrended data are in grey, and the model and 16\%–84\% confidence levels are in blue (for TESS transits) and red (for ground-based transits). In each upper plot, the point at x=-0.05 represents the median uncertainty in the photometric data.}
    \label{fig:LCs_fitted}
\end{figure*}

\begin{figure*}
    \centering
    \includegraphics[width=1.03\textwidth]{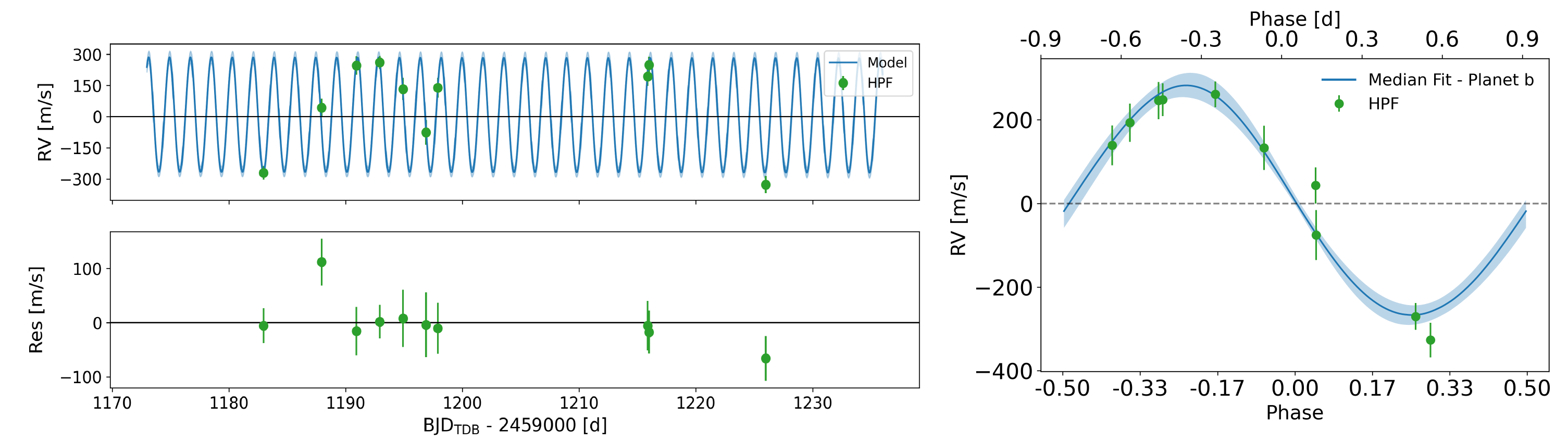}
    \caption{{\it Left}: Time series of the HPF RV data and their residuals after subtracting the best-fitting model. The best-fitting model derived from the RV-photometry joint fit is plotted in blue and the 16\%–84\% confidence interval in light blue. We show data binned by night in green. Residuals of the fit are plotted in the bottom plot below the time series.\\
    {\it Right}: Phase-folded HPF RV curve of TOI-6383.}
    \label{fig:RV_phasefolded}
\end{figure*}

We perform a joint fitting of the reduced two TESS and six ground-based transit light curves and RV data using the \texttt{Python} package \texttt{exoplanet} (\citealp{ForemanMackey2021}). It utilizes \texttt{PyMC3}, a Hamiltonian Monte Carlo (HMC) posterior sampling algorithm (\citealp{Salvatier2016}). \\
\indent Each photometry dataset is fitted with separate quadratic analytic limb darkening coefficients, using the parameterization of \cite{Kipping2013}. Additionally, we incorporate a white noise model by including a jitter term for each ground-based photometry dataset. In the case of the TESS photometry, we include a dilution term to account for the presence of blended or spatially unresolved nearby background stars. The dilution term is adjusted individually for each TESS sector, as the positioning of the target and background stars on the camera pixels varies. In contrast, given the PSF and aperture size are much smaller than the separation from the companion, we assume that the ground-based photometry from the RBO transits remains uncontaminated. Moreover, the likelihood function for the TESS photometry includes a Gaussian Process (GP) kernel to model the quasiperiodic signal which is likely an artefact from the photometry reduction of the FFI (see \citealp{Kanodia2022} for more details). The separate rotation kernels result in a recurrence timescale of 6.37$^{+6.37}_{-4.49}$ days in Sector 19 dataset and 11.63$^{+1.97}_{-2.33}$ days in Sector 59 dataset. \\
\indent The RV curve is fitted with a Keplerian model, allowing the eccentricity to vary. Additionally, we incorporate an RV offset and jitter term specific to HPF, as well as a linear RV trend to accommodate long-term drifts, encompassing both instrumental and astrophysical factors, such as the presence of an additional planetary companion.
\indent Convergence is checked using the Gelman–Rubin statistic, satisfying $\hat{R} \leq$ 1.1 (\citealp{GelmanRubin1992,Ford2006}). The planetary parameters derived from the joint fit are shown in Table~\ref{tab:planetary_parameters}. The planet is on a 1.791-day orbit around TOI-6383A, at a scaled distance of a / R$_\star$ = 10.86$\pm$0.32. It has a mass of 1.040 $\pm$ 0.094 
M$_J$ and a radius of 
1.008$^{+0.036}_{-0.033}$~R$_J$. The fitted TESS and ground-based light curves are in Figure \ref{fig:LCs_fitted}, while the phase-folded RV curve can be found in Figure \ref{fig:RV_phasefolded}, both along with the residuals from the fit.\\

\noindent To confirm these results, we also performed a joint analysis on the RV data and the TESS and ground-based transits with the \texttt{idl} script Transit Light Curve Modeller\footnote{\url{http://www.transits.hu/}} (\texttt{TLCM}; \citealp{Csizmadia2020}) to have an independent check of the results of the planet and system parameters. The photometric noise model is based on a wavelet model by \cite{CarterWinn2009}, extended by \cite{Csizmadia2023}. For this specific analysis, it has been implemented to model multiple band transit observations. All parameters fitted with the two aforementioned scripts agree within 2-$\sigma$, confirming the previous results.\\

\begin{table*}[]
    \centering
    \begin{tabular}{l l c }
\hline \hline
    Parameter & Symbol, units & Value \\
    \hline
\textbf{Orbital parameters}\\
Orbital period & P \hspace{1.0cm}[d] & 1.79058695$^{+{8.7e-7}}_{-{8.6e-7}}$ \\
Eccentricity$^\dagger$ & e & 0.050$^{+0.042}_{-0.033}$, $^{+0.086}_{-0.047}$ \\
Periastron angle & $\omega$ \hspace{0.9cm} & 0.42$^{+1.26}_{-1.08}$ rad = 23.8$^{+72}_{-62}$$^\circ$ \\
Semi-major axis & a \hspace{1.0cm} [AU] & 0.02292 $\pm$ 0.00037\\
RV Semi-amplitude & K \hspace{1.0cm}[m s$^{-1}$] & 277 $\pm$ 23 \\
RV offset velocity$^a$ & $\gamma$ \hspace{1.05cm}[m s$^{-1}$] & -81 $\pm$ 19 \\
RV trend & $\dot{\gamma}$ \hspace{0.98cm} [m s$^{-1}$ yr$^{-1}$] & -30 $\pm$ 97\\
RV jitter & $\sigma_{RV}$ \hspace{0.6cm} [m s$^{-1}$] & 31$^{+26}_{-21}$ \\
Impact parameter & b & 0.22$^{+0.12}_{-0.15}$ \\
Orbital inclination & i \hspace{1.05cm} [$^\circ$] & 88.80$^{+0.78}_{-0.68}$ \\
\\
\textbf{Transit parameters}\\
Transit mid-time & T$_c$ \hspace{0.9cm}[BJD$_\mathrm{TDB}$] & 2459933.59545$^{+{2.0e-4}}_{-{2.1e-4}}$ \\
Transit duration & T$_{14}$ \hspace{0.75cm}[d]& 0.0641$^{+0.0020}_{-0.0017}$ \\
Scaled radius & R$_p$ / R$_\star$ & 0.2245$^{+0.0042}_{-0.0041}$ \\
Scaled semi-major axis & a / R$_\star$ & 10.68$\pm$0.32\\
Photometric jitter$^b$ & $\sigma_\mathrm{TESS 19}$ \hspace{0.175cm}[ppm] & 6524$^{+160}_{-150}$ \\
& $\sigma_\mathrm{TESS 59}$ \hspace{0.175cm}[ppm] & 14901$^{+109}_{-107}$  \\
& $\sigma_\mathrm{RBO 2108}$ \hspace{0.0cm}[ppm] & 8432$^{+1505}_{-1420}$ \\
& $\sigma_\mathrm{RBO 3008}$ \hspace{0.0cm}[ppm] & 47$^{+650}_{-44}$ \\
& $\sigma_\mathrm{RBO 0809}$ \hspace{0.0cm}[ppm] & 72$^{+1047}_{-68}$ \\
& $\sigma_\mathrm{RBO 2609}$ \hspace{0.0cm}[ppm] & 52$^{+723}_{-48}$ \\
& $\sigma_\mathrm{RBO 3010}$ \hspace{0.0cm}[ppm] & 46$^{+574}_{-43}$\\
& $\sigma_\mathrm{RBO 1511}$ \hspace{0.0cm}[ppm] & 91$^{+1762}_{-87}$ \\
Dilution$^{c,d}$ & D$_\mathrm{TESS 19}$ & 0.882$^{+0.042}_{-0.040}$ \\
& D$_\mathrm{TESS 59}$ & 0.894$\pm$0.032 \\
\\
\textbf{Planetary parameters}\\
Mass & M$_p$ \hspace{0.7cm} [M$_J$] & 1.040 $\pm$ 0.094 \\
& M$_p$ \hspace{0.7cm} [M$_\Earth$] & 331 $\pm$ 30 \\
Radius & R$_p$ \hspace{0.8cm} [R$_J$] & 1.008$^{+0.036}_{-0.033}$ \\
& R$_p$ \hspace{0.89cm}[R$_\Earth$] & 11.29$^{+0.41}_{-0.37}$ \\
Density & $\rho_p$ \hspace{0.95cm}[g cm$^{-3}$] & 1.26$^{+0.18}_{-0.17}$ \\
Planetary insolation & S \hspace{1cm} [S$_\Earth$]& 51.3 $\pm$ 6.3 \\
Equilibrium temperature$^e$ & T$_{eq}$ \hspace{0.85cm}[K] & 745 $\pm$ 23\\
\hline
\hspace{0.3cm}
\end{tabular}
\caption{Orbital, transit and planetary parameters based on the \texttt{exoplanet} package solution.\\
The reported errorbars in the third and fourth columns correspond to 1$\sigma$ uncertainty, while
$^\dagger$ both 1$\sigma$ and 2$\sigma$ confidence levels are reported for the eccentricity.\\
$^a$ In addition to the absolute RV in Table \ref{tab:stellar_parameters}.\\
$^b$ Jitter added in quadrature to photometric instrument error separately for each transit. \\
$^c$ Dilution due to the presence of background stars in TESS aperture not accounted for in the \texttt{eleanor} flux. \\
incident flux. \\
$^d$ We assume the planet to be a blackbody with zero albedo and perfect energy redistribution to estimate the equilibrium temperature.\\
}
    \label{tab:planetary_parameters}
\end{table*}

\begin{figure}
    \centering
    \includegraphics[width=0.48\textwidth]{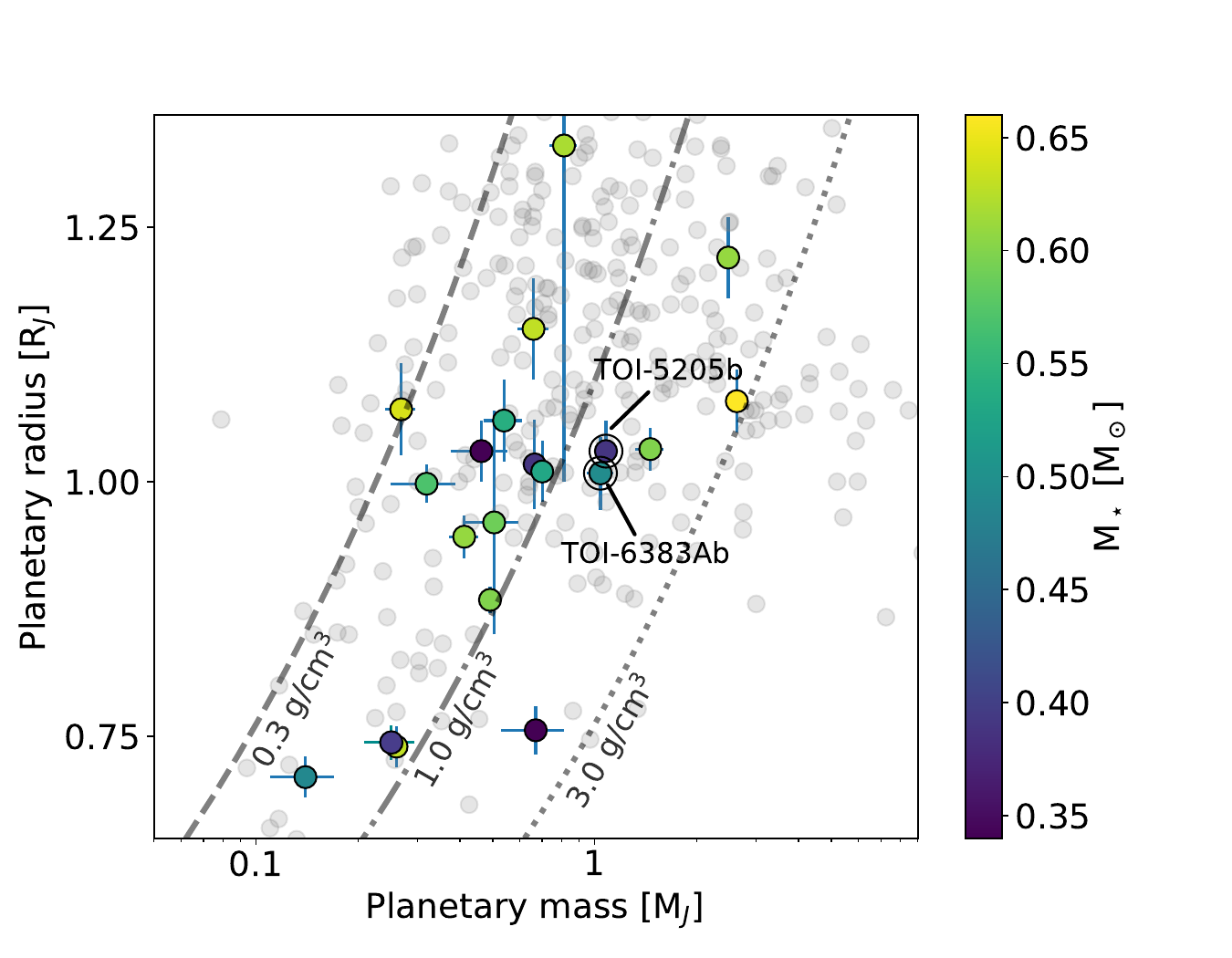}
    \includegraphics[width=0.48\textwidth]{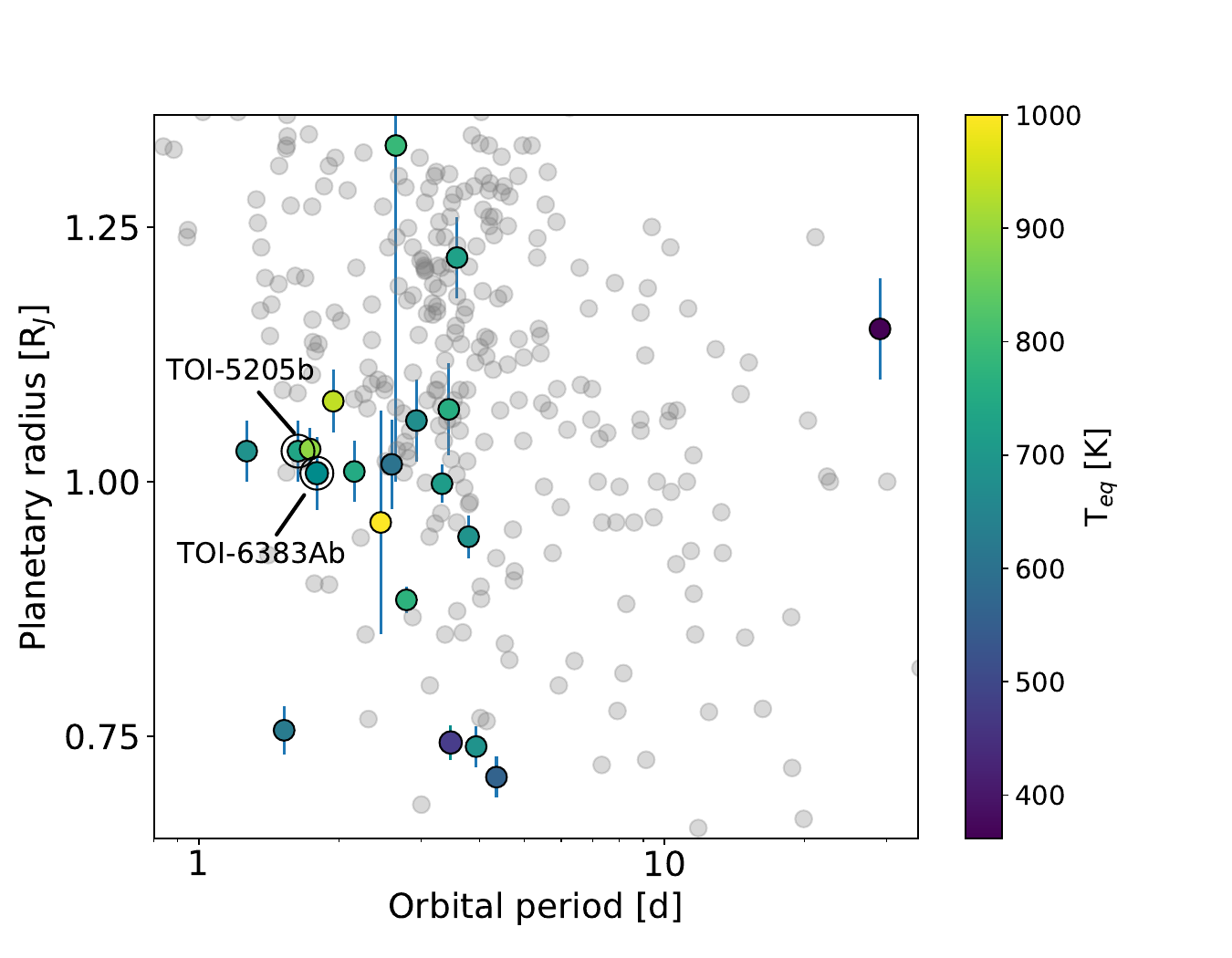}
    \caption{{\it Top plot}: Mass–radius plane of all known transiting GEMS (coloured in the foreground, with errorbars) and known planets around F, G, and K stars (in transparency in the background), with planets coloured by host star mass. TOI-6383Ab and the similar TOI-5205 b are circled. We also plot the density contours for 0.3 g cm$^{-3}$, 1 g cm$^{-3}$ and 3 g cm$^{-3}$. NGTS-1 b has an imprecise estimate of the radius (the large errorbar is clearly visible in the top part of the plot) to due its grazing transit.\\
    {\it Bottom plot}: orbital period vs. planet radius, coloured by planetary equilibrium temperature. TOI-1899b is highlighted because of its long orbital period ($\sim$ 29 days) with respect to the other GEMS.\\ 
    }
    \label{fig:diagrams}
\end{figure}

\section{Discussion}
\label{sec:discussion}
\subsection{TOI-6383Ab in context}
TOI-6383Ab joins the short but growing list of GEMS. Due to the small sample size of confirmed GEMS, it is still hard to make robust conclusions regarding trends in their planetary and orbital parameters. This planet is found as part of the {\it Searching for GEMS survey} and allows us to delve deeper into the statistics of the $\sim$25 transiting GEMS discovered up to this date (see the complete list in Table 1 in \citealp{Kanodia2024}).\\

\noindent TOI-6383Ab is a Jupiter-like giant planet with a radius of 1.008$^{+0.036}_{-0.033}$~R$_J$, a mass of 1.040 $\pm$ 0.094 M$_J$ and a mean density of 1.26$^{+0.18}_{-0.17}$ g cm$^{-3}$. Using the NASA Exoplanet Archive (\citealp{Akeson2013}) database, in Figure \ref{fig:diagrams} we put TOI-6383Ab in context by plotting it in parameters spaces together with all known GEMS and giant planets around F, G and K-type stars. Chosen planets as GEMS are in the following range of parameters: stellar mass [0.30, 0.65] M$_\Sun$ and effective temperature [3000, 4000] K and planetary mass [40,900] M$_\Earth$ and radius $>$8 R$_\Earth$.\\ 
\indent TOI-6383Ab is similar to TOI-5205 b (\citealp{Kanodia2023}; both highlighted in the two diagrams in \ref{fig:diagrams}) in terms of planetary mass (1.04 M$_J$ and 1.08 M$_J$, respectively) and radii (1.01 R$_J$ and 1.03 R$_J$, respectively) \textemdash~and consequently mean density \textemdash~and equilibrium temperature (745 K and 737 K, respectively).\\ 
\indent The upper panel in Figure \ref{fig:diagrams} shows the relation between planetary mass and planetary radius, coloured by stellar mass. 0.3 g cm$^{-3}$, 1 g cm$^{-3}$ and 3 g cm$^{-3}$ density contours are also plotted for a first look into planet type classification and interior structure.\\
\indent In the lower plot, we show planetary mass and orbital period of GEMS and FGK giant planets, coloured by planetary equilibrium temperature. The majority of transiting GEMS have an orbital period below 4.5 days. However, GEMS are cooler than FGK hot Jupiters (see colour scale of the lower plot, with a maximum of 1000 K) due to the cooler and less massive M-dwarf hosts. Therefore, they are unlikely to be inflated due to ohmic dissipation caused by high stellar insolation (\citealp{ThorngrenFortney2018}). \\

\noindent Among the $\sim$20 confirmed transiting GEMS so far, this is the tenth whose star has a bound stellar companion (see \cite{Jordan2022} for HATS-74Ab, \citealp{Canas2023} for TOI-3984Ab and TOI-5293Ab, \citealp{Canas2022} for TOI-3714b, \citealp{Hartman2024} for TOI-762 Ab and Reji et al., in preparation for TOI-5688Ab). Details on these planets and their host stars are reported in Table \ref{tab:binaries}. The distribution of the M-dwarf binary system projected separation peaks at $\sim$20-40 AU (see \citealp{Winters2019,Susemiehl2022}), but no GEMS are found in systems with such tight systems. This could imply that close binary companions might disrupt planet formation processes, while wider separations ($\gtrsim$ 200 AU) are more favourable for the development of giant planets around M-dwarfs.\\
\begin{table*}[]
    \centering
    \begin{tabular}{l c c c c c}
    \hline \hline
      Planet name & Star mass A & Star mass B & Projected distance & Planet mass & Reference \\ 
      & [M$_\Sun$] & [M$_\Sun$] & [AU] & [M$_J$] & \\
    \hline
      HATS-74Ab & 0.6010 $\pm$ 0.0080 & 0.2284 $\pm$ 0.0078 & 238 & 1.46 $\pm$ 0.14 & \cite{Jordan2022} \\
      TOI-3714b$^{\dagger 1}$ & 0.53 $\pm$ 0.02 & ... & 302 & 0.70 $\pm$ 0.03 & \cite{Canas2022}\\
      TOI-3984Ab$^{\dagger 1}$ & 0.49 $\pm$ 0.02 & ... & 356 & 0.14 $\pm$ 0.03 & \cite{Canas2023}\\
      TOI-5293Ab & 0.54 $\pm$ 0.02 & ... & 579 & 0.54 $\pm$ 0.07 & \cite{Canas2023}\\ 
     TOI-762Ab & 0.442 $\pm$ 0.025 & 0.227 $\pm$ 0.010 & 319 & 0.251 $\pm$ 0.042 & \cite{Hartman2024} \\
      TOI-5634Ab & 0.556 $\pm$ 0.022 & ... & 1230 & 0.58$^{+0.41}_{-0.35}$$^{\dagger 3}$ & \cite{Kanodia2024} \\
      K2-419Ab & 0.562 $\pm$ 0.024 & ... & 520 & 0.617 $\pm$ 0.047 & \cite{Kanodia2024} \\
      TOI-6034b$^{\dagger 2}$ & 0.514$^{+0.025}_{-0.022}$ & ... & 4700 & 0.798 $\pm$ 0.075 & \cite{Kanodia2024}\\
      TOI-5688Ab & 0.64 $\pm$ 0.06 & 0.29 $\pm$ 0.06 & 1100 & 0.34 $\pm$ 0.15 & Reji et al., submitted\\
      TOI-6383Ab & 0.458 $\pm$ 0.011 & 0.187 $^{+0.011}_{-0.013}$ & 3126 & 1.040 $\pm$ 0.094 & This work \\
      \hline
      \hspace{0.3cm}
    \end{tabular} 
    \caption{Details on host star, its stellar companion and the hosted planet for the ten GEMS hosted by a binary system.\\
    $^{\dagger 1}$ The stellar companion is a resolved white dwarf. Their stellar mass is not reported in this Table because, as mentioned in the corresponding paper, it was calculated using models, while ideally it should be estimated using low-resolution optical spectra.\\
    $^{\dagger 2}$ TOI-6034b is the first GEMS host which is part of a wide-separation binary with a main sequence companion (late-F star).\\
    $^{\dagger 3}$ A mass upper limit is given in the corresponding paper.}
    \label{tab:binaries}
\end{table*}

\begin{figure}
    \centering
    \includegraphics[width=0.48\textwidth]{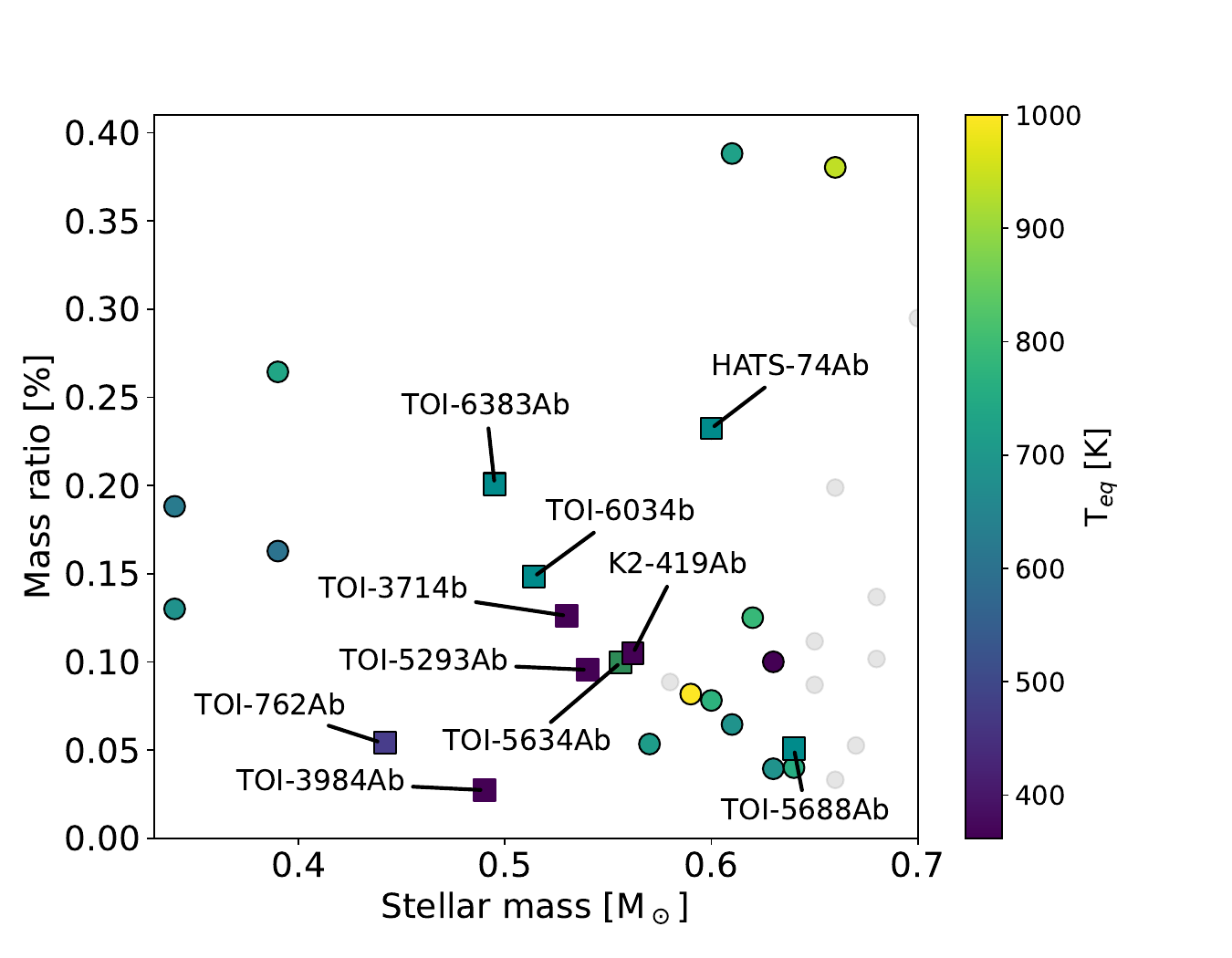}
    \caption{Transiting GEMS (coloured) and FGK (in grey) stellar mass vs the planet-to-star mass ratio, coloured by planetary equilibrium temperature. GEMS discovered by the {\it Searching for GEMS survey} are circled in black. GEMS in a stellar binary system are represented with a squared marker and labelled with their name. TOI-4201b and HATS-76b correspond to the two higher mass ratios in the plot, clearly visible on the top right.}
    \label{fig:mass_ratio}
\end{figure}
All discovered GEMS can be seen in Figure \ref{fig:diagrams}, as well as in Figure \ref{fig:mass_ratio}. Here GEMS and exoplanets around FGK stars are plotted showing the stellar mass vs the planetary-to-stellar mass ratio, coloured by planetary equilibrium temperature. GEMS discovered and published by our Survey are highlighted with a black circle.\\
\indent The typical large planet-to-star mass ratio ($\sim$0.2\%) of such systems 
may imply strong tidal interactions between the two bodies, in the case of short orbital periods (see for example the Equations for tidal interactions in \citealp{Kopal1978}). This could open a new door to the study of gravitational interactions and their consequences on the planetary shape and changes of the orbital elements. Moreover, \cite{Zanazzi2024} showed that in the case of cool stars ($T_{eff} \lesssim$ 6100 K, with radiative cores) resonance locking can significantly damp the stellar obliquity, orbital semimajor axis, and eccentricity, explaining why cooler stars tend to have spin axes aligned with their hot Jupiters, unlike hotter stars, which lack radiative cores and therefore do not experience the same degree of tidal evolution. In the case of an early M-dwarf such as TOI-6383A, the obliquity can be damped even slower due to the very large convective envelope. This makes GEMS interesting targets for Rossiter-McLaughlin measurements to test tidal theories.\\

\noindent Due to their scaled distance to the star (typically, a/$R_\star \gtrsim$ 10), these objects may not be tidally locked or circularized like hot Jupiters yet. According to \cite{GoldreichSoter1966}, Equation (3) of \cite{Waalkes2021}, \cite{Jackson2008} and \cite{Persson2019}, the circularization timescale for TOI-6383Ab is $10^9-10^{10}$ yrs depending on the model and assuming Jupiter's tidal quality factor Q=10$^5$ for the planet, and the tidal decay timescale is $10^{11}-10^{12}$ yrs. This suggests that tidal interactions on these planets are minimal and likely insufficient to cause complete circularization within the star's lifetime. Consequently, the interior and atmospheric properties of GEMS may differ significantly from those of hot Jupiters, which experience more intense tidal interactions. In conclusion, despite the limited number of confirmed GEMS, we can already hypothesize that their interior, atmospheric and orbital characteristics might not be as similar to hot Jupiters as they might seem at first glance.\\
\subsection{TOI-6383Ab's interior}
\label{subsec:evol_tracks}
\noindent We provide a simple calculation of the heavy element content in the planet. From the relation provided by \cite{Thorngren2016} (fit of their Figure 7), the mass of heavy elements results in $\sim$ 59 M$_\Earth$, with a scatter of about 10 M$_\Earth$ due to the scatter of the fit and the uncertainty in the planetary mass. We note that some caveats of such a relation must be considered. The intrinsic spread in the data suggests variability that may not be fully captured by the model, potentially influenced by factors such as the planet's migration history and stochastic nature of formation. Additionally, observational uncertainties and the simplifications in the model, such as not accounting for atmospheric effects or composition gradients, may contribute to the observed scatter in heavy element mass estimates. 
Assuming runaway gas accretion on to a $\sim$ 10 M$_\Earth$ core, the remaining $\sim$ 50 M$_\Earth$ of heavy-elements are likely to be spread out through the H/He envelope, similarly to TOI-5205 b (\citealp{Kanodia2023}).\\

\begin{figure}
    \centering
    \includegraphics[width=0.48\textwidth]{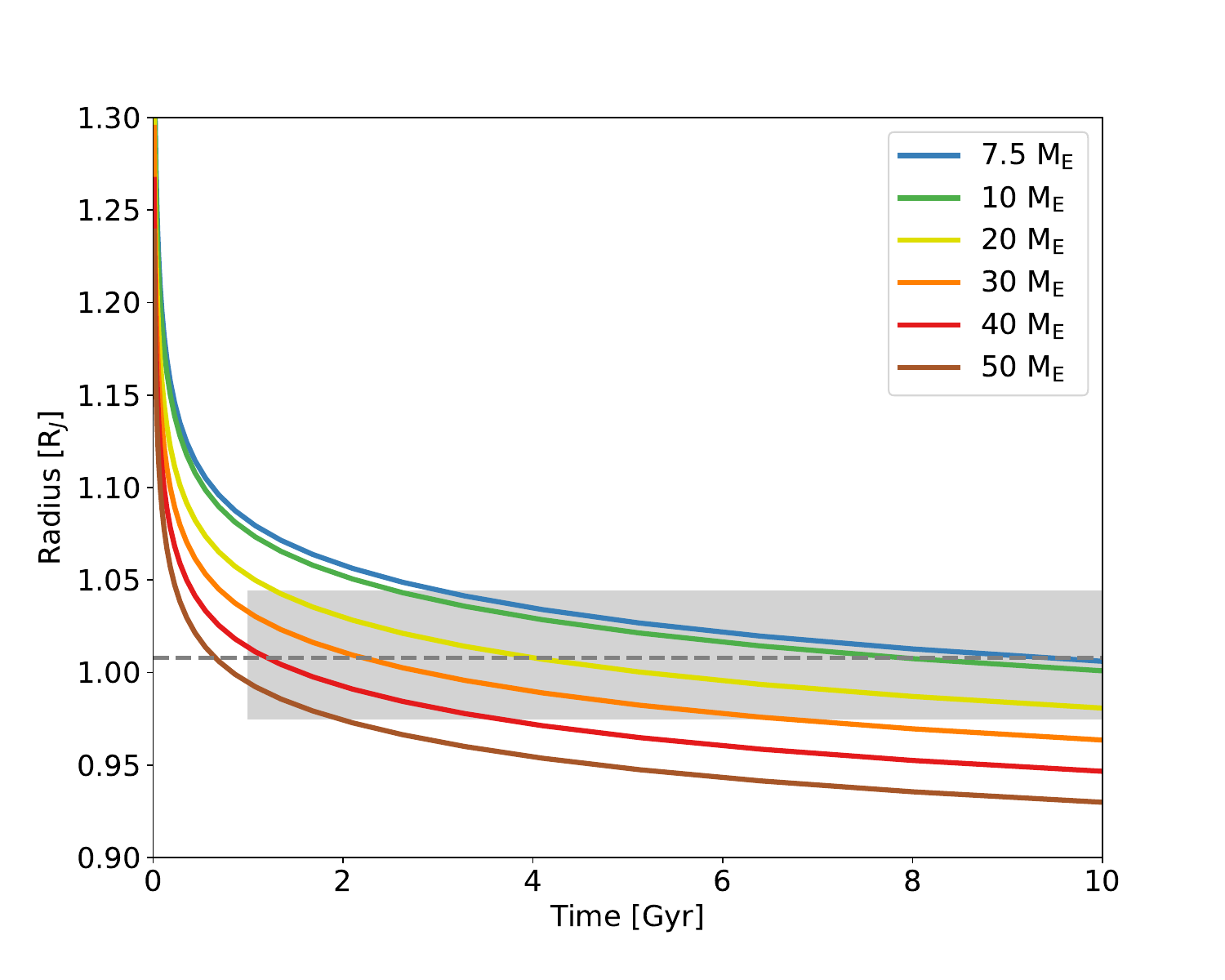}
    \includegraphics[width=0.48\textwidth]{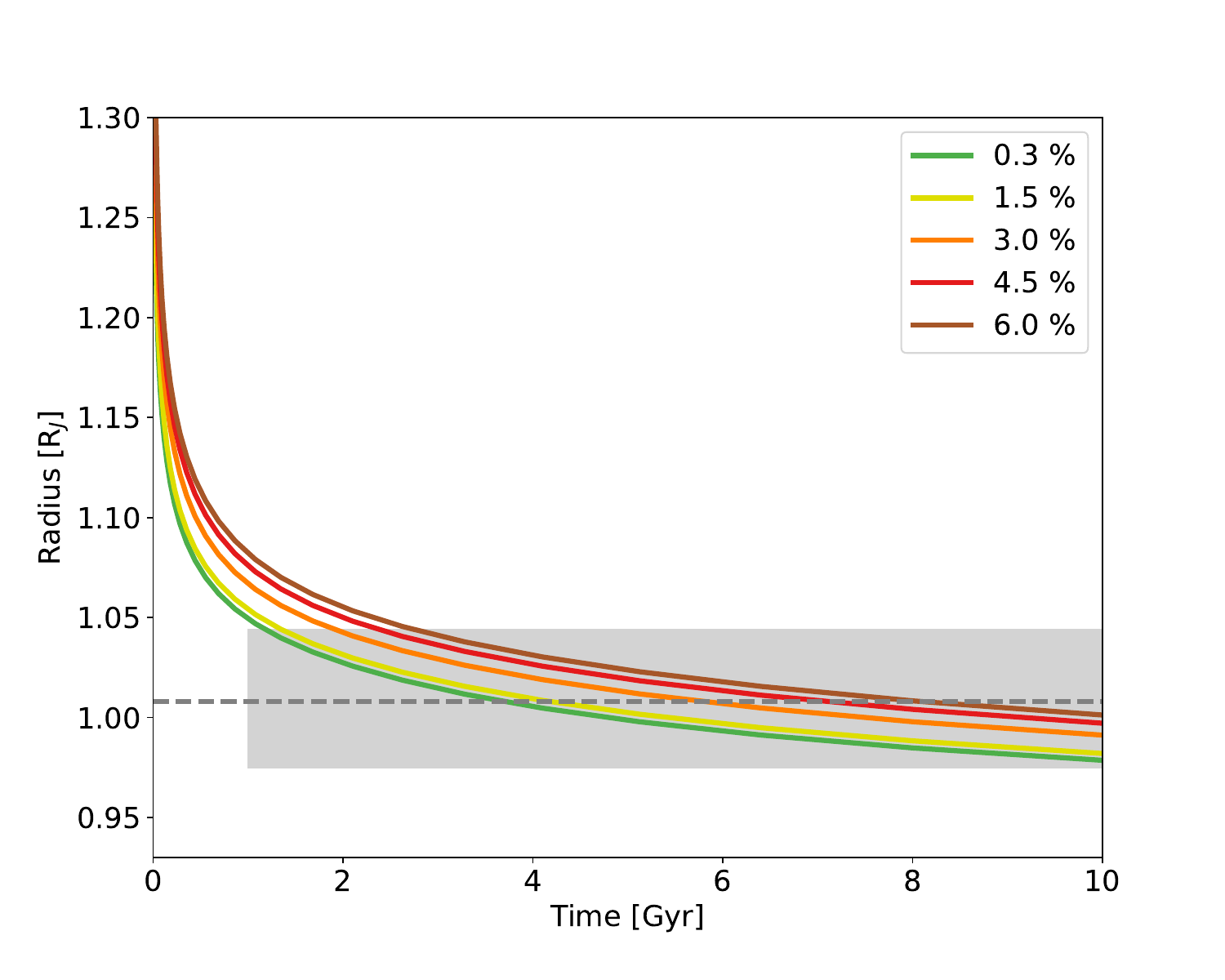}
    \caption{The radius evolution in time of a planet with the mass of TOI-6383Ab, according to the model of \cite{MullerHelled2021}.\\
    {\it Top plot}: Evolution track of the planetary radius varying its heavy-element (core) mass, and keeping the atmospheric metallicity fixed at the TOI-6383A value. The horizontal dashed line indicates the measured radius of the planet. The grey region corresponds to the 16$^\mathrm{th}$ and 84$^\mathrm{th}$ quantiles on the planetary radius, while for the age we only have an indication that the star is not young ($>$1 Gyr). Colours indicate the core mass, which ranges between 7.5 and 50 M$_E$. \\
    {\it Bottom plot}: Evolution track of the planetary radius varying the planetary atmospheric metallicity from 0.3\% to a maximum of 6\% as indicated by the colours (with a fixed core heavy-element mass of 20 M$_\Earth$).}
    \label{fig:evol_track_radius}
\end{figure}

\noindent We utilized the planetary evolution model implemented in the \texttt{Python} package \texttt{planetsynth}, developed by \cite{MullerHelled2021}. This model is designed to simulate the thermal and structural evolution of giant planets by solving the 1D hydrostatic equilibrium equations (\citealp{Kippenhahn2013}). Key inputs for the model include planetary mass, overall composition (including atmospheric metallicity), and the intensity of stellar radiation. The model assumes that giant planets form via the core accretion process and initially cool adiabatically within a "hot start" scenario (\citealp{Marley2007}), characterized by a large initial radius. The evolution of the planet is then tracked over time, providing predictions for changes in planetary radius, luminosity, effective temperature, and surface gravity.\\
\indent A core-envelope structure is assumed, where the majority of heavy elements are concentrated in a central, compact core. It also takes into account the effects of stellar irradiation on the planetary atmosphere, along with the enrichment of heavy elements in the atmospheric opacity (\citealp{Burrows2007,Mueller2020}).\\
\indent For TOI-6383Ab, we computed the evolutionary track of the planet's radius by fixing the planetary mass and the incident stellar irradiation based on values derived from a joint fit (see Table \ref{tab:planetary_parameters}). We then varied the bulk metallicity and atmospheric content of metals to explore their impact on the cooling process. In the top plot of Figure \ref{fig:evol_track_radius}, we show the effect of varying the core mass from 7.5 to 50 M$\Earth$ while keeping the atmospheric metallicity Z$_\mathrm{atm}$ fixed at the solar value of 0.01. According to such a plot, the concentration of heavy elements in the core is below  $\sim${\bf40}~M$_\Earth$, and more likely $\sim${\bf 7.5--30}~M$_\Earth$, which aligns with the prediction made with \cite{Thorngren2016}'s results. Due to the uncertainty in the stellar metallicity, we also re-calculated the evolutionary track with Z$_\mathrm{atm}$ fixed at higher values to account for the fact that TOI-6383A seems to be a moderately metal-rich star. However, the result does not change.\\
\indent Despite the robustness of the \texttt{planetsynth} model, several caveats must be acknowledged. First, the assumption of a "hot start" scenario introduces uncertainty, as the initial conditions of giant planets can vary significantly, particularly if non-adiabatic processes are involved. The assumption of adiabatic cooling may not be valid for all planets, especially those with internal mixing or large composition gradients (\citealp{Vazan2013,Vazan2015}). Additionally, the model does not account for clouds or grains in the atmosphere, which could impact cooling rates by trapping heat (\citealp{Vazan2013,Poser2019}). The simplified treatment of stellar irradiation may also lead to inaccuracies, particularly for not extremely irradiated planets (\citealp{Valsecchi2015}). Finally, the assumption of a core-envelope structure without considering composition gradients or more complex internal structures, such as extended, dilute cores (see e.g \citealp{Lozovsky2017}) could limit the accuracy of the model predictions.\\
\indent Comparing the result from such a model with the fitted value of the planetary radius (see Table \ref{tab:planetary_parameters} and the grey-shaded region in the top plot of the figure, assuming the system is not younger than 1 Gyr), we can conclude that the planetary radius is not inflated, as expected from its equilibrium temperature. 
In the bottom plot, we fixed the mass of the core at 20 M$_\Earth$ and varied the atmospheric metallicity from 1 to 20 M$_\Earth$ (or equivalently from 0.3\% to 6\%). No value is particularly preferred in this case since all curves agree with our estimated planetary radius.\\

\noindent However, given the assumptions and uncertainties inherent in the actual interior structure models, particularly when only some planetary parameters - e.g. mass, radius and stellar irradiation - are known, our aim is not to draw definitive conclusions but rather to assess the plausibility of different interior compositions within the context of our observational constraints.


\subsection{TOI-6383Ab's formation}
In this Section, we will consider both actual giant planet formation scenarios - CA and GI - to assess if one is preferred over the other for TOI-6383Ab.\\
\indent \cite{Kanodia2024} widely discussed the challenges of forming GEMS. First of all, there may not be enough dust mass in protoplanetary discs to support the formation of GEMS during the protoplanetary phase via core accretion. As a matter of fact, due to the low mass of M-dwarfs \textemdash~ranging from $\sim$0.08$M_\Sun$ to $\sim$0.65$M_\Sun$ \textemdash~the median Class II disc around such a star is also expected to have a lower median mass compared to solar-type stars (\citealp{Andrews2013, Pascucci2016}). The second issue that should be pointed out is the formation timescale of a solid core massive enough to initiate the runaway gaseous accretion phase (\citealp{Laughlin2004,IdaLin2005}). At the same orbital distance, the Keplerian orbital timescale is longer for bodies orbiting M-dwarfs with respect to FGK stars. Forming a 10 M$_\Earth$ core around an M-dwarf star before the disc dissipates is a challenge (see \citealp{Laughlin2004} and reference therein).\\

\noindent To address the possibility that TOI-8383Ab formed via CA, we make a simple estimate of the mass budget available in the protoplanetary disc to form the planet via core accretion can be done with some assumptions. First, it is reasonable to assume that the disc \textemdash~and therefore the mass that it contains \textemdash~should not be truncated by the companion star through gravitational interaction due to the high physical distance between the two stars, unless in the presence of a highly eccentric planetary orbit. The typical size of a protoplanetary disc is a few tens of AU in the sub-millimeter/millimeter dust up to a few hundreds for the extension of the gas discs (\citealp{Barenfeld2017,Tazzari2017,Ansdell2018}), while the projected distance between the two stars is $\sim$3100 AU (see Table \ref{tab:stellar_parameters}). \\
\indent We can assume the disc-to-star mass ratio between 0.2 and 0.6\% (\cite{Andrews2013}, in the Taurus forming region). This implies that the total mass of the protoplanetary disc of TOI-6383A was between $\sim$ 300 and 900 M$_\Earth$, respectively when assuming 0.2 and 0.6\% for the disc-to-star mass ratio. With a canonical gas-to-dust mass content ratio of the interstellar medium (ISM) of 100:1, the dust mass ranges between $\sim$3 and 9 M$_\Earth$. Even supposing that the entire available mass converged to form TOI-6383Ab, this quantity would be insufficient to form a core of 10 M$_\Earth$ in the core accretion formation scenario, necessary to initiate runaway gaseous accretion. Moreover, the content of dust available is definitely lower than the predicted mass from Section \ref{subsec:evol_tracks}, of 7.5-30 M$_\Earth$ in the core and an additional few M$_\Earth$ in the atmosphere.\\
\indent Even in the most massive disc (i.e., almost gravitational unstable, $\sim$ 150 -- 200 M$_\Earth$), a planet of $\sim$60 M$_\Earth$ of heavy elements (contained partly in the core and partly in the atmosphere, as calculated in Section \ref{subsec:evol_tracks}) would require a formation efficiency of $\sim$ 30 -- 40\% to form via core accretion, in contrast with the nominal value of 10\% (\citealp{Liu2018}) from formation models of pebble accretion. \\
\indent However, there is some scatter in these scaling relations. We could hypothesize that the disc around TOI-6383A was more massive than predicted. \cite{Andrews2013} pointed out some outliers within their sample of Taurus discs, noting their unusually high masses, whose disc-to-stellar mass ratio is around 10\%. If that is the case, there could have been enough material to form the core and then subsequently the planet. On the other hand, with different assumptions on the gas-to-dust ratio (e.g. $\sim$70:1 for the solar medium, \citealp{Bohlin1978}), the total mass content can vary between $\sim$4 and $\sim$13 M$_\Earth$ (calculated assuming 0.2 and 0.6\% for the disc-to-star mass ratio, respectively).\\
\indent We also want to highlight a few caveats in this discussion. First of all, the dust mass measured by the Atacama Large Millimeter/submillimeter Array (ALMA) in mm-sized particles does not accurately reflect the original mass budget. The formation process has already begun, therefore part of the dust is already aggregated in larger particles than the observed mm-sized ones (\citealp{Greaves2010, NajitaKenyon2014}). Moreover, due to interactions with the gas present in the disc and the effect of radiation pressure, dust particles tend to migrate radially inward in the disc, eventually falling onto the central star. This phenomenon also causes dust depletion (\citealp{Appelgren2023}). Furthermore, the dust mass in protoplanetary discs is typically estimated based on continuum flux measurement of mm-size dust particles by assuming blackbody emission and with a single wavelength, at 850 $\mu$m (\citealp{Hildebrand1983}). The straightforward relation between flux and dust mass is valid if there are no structures in the disc \textemdash~such as rings or gaps, \cite{Liu2022} \textemdash~and if the continuum emission is optically thin (\citealp{Hildebrand1983}). If one of these conditions is not met, the amount of dust can be underestimated by a factor between 3 and 10 (\citealp{Liu2022}).\\
\indent It is worth mentioning that recent numerical simulations by \cite{Savvidou2024} of planet formation via pebble and gas accretion in a viscously evolving protoplanetary disc, suggest that, especially if the planet forms in the inner disc, hidden dust mass, coupled with early planet formation, might address the hypothetical mass budget problem.\\

\noindent Given the limitations of existing dust mass measurements from ALMA in accessing the primordial mass budget available for formation via core accretion \cite{Miotello2023}, we instead consider the most massive discs possible before the onset of gravitational instability (\citealp{Boss2006, Boss2011, BossKanodia2023}). Gravitational instability becomes possible for discs that are $\sim$ 10 -- 15 \% the host star mass for $\sim$0.46 $M_\Sun$ stars (Figure 7 in \citealp{BossKanodia2023}). Assuming this to be the maximum disc mass for core accretion, and the standard gas-to-dust mass ratio of 100:1, this corresponds to a minimum dust mass content of $\sim$ 150 -- 200 M$_\Earth$. A compilation of ALMA observations (Figure 6 of \citealp{Manara2023}) also shows that the most massive disc around an M-dwarf has $\sim$ 140 M$_\Earth$ of dust mass, below the minimum amount required to initiate gravitational instability. \\

\noindent In summary, while the CA scenario could explain the formation of TOI-6383Ab under the assumption of an unusually massive dust-rich disc and a higher-than-average core formation efficiency, this scenario still faces significant challenges. It also entails the core formation efficiency for GEMS to be much higher than 10\%, or if we assume that the core formation efficiency is still 10\%, then these GEMS accrete a significant heavy-element fraction through post-formation processes such as late-stage pollution (\citealp{Liu2015}). Besides the high demand for dust mass in the disc and a high formation efficiency, another challenge in core accretion is the pebble isolation mass (\citealp{Liu2019}). The conclusion of the pebble accretion phase is significantly influenced by the stellar mass and defines the characteristic mass of planets, as it effectively stops further growth through pebble accretion.\\
\indent Alternatively, GI is shown to be a plausible alternative by recent studies (see e.g \citealp{BossKanodia2023}), especially given the possibility of forming the planet in a marginally stable disc. However, in the case of TOI-6383A, gravitational instability would require a minimum dust mass greater than any observed in discs around M-dwarfs.\\
\indent Currently, neither scenario can be definitively identified as the formation path for TOI-6383Ab. The conventional mechanisms for gas giant formation are especially challenging when applied to GEMS, indicating that these are likely rare objects, regardless of whether they formed through CA or GI, in agreement with  transiting occurrence rate studies from \cite{Gan2023} and \cite{Bryant2023}. 
A larger sample of transiting GEMS with well-characterized planetary properties and stellar metallicities (despite the complexities in its determination in the case of M-dwarf) could help distinguish between the two, as is the goal of our \textit{Searching for GEMS survey} (\citealp{Kanodia2024}).

\section{Conclusions}
We report the discovery of TOI-6383Ab, a massive Jovian planet around a nearby M star as part of the {\it Search for Giant Exoplanets around M-dwarf Stars} survey. TOI-6383Ab has radius of 1.008 $^{+0.036}_{-0.033}$ R$_J$, mass 1.040 $\pm$ 0.094 M$_J$, and density 1.26 $^{+0.18}_{-0.17}$ g cm$^{-3}$ around the early-type M-dwarf star TOI-6383A. The target was identified in TESS data, and then ground-based transit and radial velocity observations followed to confirm and characterize it.\\
\indent We also detected a 0.2 $M_\sun$ M-dwarf stellar companion of TOI-6383A. The discovery of TOI-6383Ab marks the seventh instance among $\sim$25 confirmed transiting GEMS where the host star has a bound stellar companion. \\
\indent As discussed, the formation of GEMS presents significant challenges, particularly when considering the conventional mechanisms of gas giant formation, CA and GI. Both scenarios require specific and often rare conditions, such as unusually massive and dust-rich protoplanetary discs, to successfully form GEMS.
The formation of TOI-6383Ab still remains an open question. It likely formed in an exceptionally massive, dust-rich disc, with a core formation efficiency exceeding the nominal 10\% value, and notably, boasting a high planet-to-stellar mass ratio of $\sim$0.2\%. Alternatively, gravitational instability offers a viable explanation, given a massive enough protoplanetary disc.

Delineating and differentiating the properties of GEMS, such as their mass, orbital characteristics, and atmospheric compositions, is essential for constraining theoretical models and refining our understanding of planetary formation processes. 
Moreover, it is worth pointing out that due to the low stellar-to-planet-mass ratio, GEMS are a favourable target when it comes to investigating the tidal decay of the planetary orbit. \\


\section{Acknowledgments}
This work was supported by the European Union's Horizon Europe Framework Programme under the Marie Skłodowska-Curie Actions grant agreement No. 101086149 (EXOWORLD).\\
\indent The work was also supported by DFG Research Unit 2440: "Matter Under Planetary Interior Conditions: High Pressure, Planetary, and Plasma Physics".\\
\indent CIC acknowledges support by NASA Headquarters through an appointment to the NASA Postdoctoral Program at the Goddard Space Flight Center, administered by ORAU through a contract with NASA.\\
\indent This work is based on observations obtained with the Hobby-Eberly Telescope (HET), which is a joint project of the University of Texas at Austin, the Pennsylvania State University, Ludwig-Maximillians-Universitaet Muenchen, and Georg-August Universitaet Goettingen. The HET is named in honor of its principal benefactors, William P. Hobby and Robert E. Eberly.\\
\indent These results are based on observations obtained with the Habitable-zone Planet Finder Spectrograph on the HET. The HPF team acknowledges support from NSF grants AST-1006676, AST-1126413, AST-1310885, AST-1517592, AST-1310875, ATI-2009889, ATI-2009982, AST-2108512, AST-2108801 and the NASA Astrobiology Institute (NNA09DA76A) in the pursuit of precision radial velocities in the NIR. The HPF team also acknowledges support from the Heising-Simons Foundation via grant 2017-0494. \\ 
\indent The HET collaboration acknowledges the support and resources from the Texas Advanced Computing Center. We thank the Resident astronomers and Telescope Operators at the HET for the skillful execution of our observations with HPF. We would like to acknowledge that the HET is built on Indigenous land. Moreover, we would like to acknowledge and pay our respects to the Carrizo \& Comecrudo, Coahuiltecan, Caddo, Tonkawa, Comanche, Lipan Apache, Alabama-Coushatta, Kickapoo, Tigua Pueblo, and all the American Indian and Indigenous Peoples and communities who have been or have become a part of these lands and territories in Texas, here on Turtle Island.\\
\indent We acknowledge the Texas Advanced Computing Center (TACC) at The University of Texas at Austin for providing high performance computing, visualization, and storage resources that have contributed to the results reported within this paper.\\
\indent The Low Resolution Spectrograph 2 (LRS-2) was developed and funded by the University of Texas at Austin McDonald Observatory and Department of Astronomy and by Pennsylvania State University. We thank the Leibniz-Institut fuer Astrophysik Potsdam (AIP) and the Institut fuer Astrophysik Goeottingen (IAG) for their contributions to the construction of the integral field units. \\
\indent Some of the observations in this paper made use of the NN-EXPLORE Exoplanet and Stellar Speckle Imager (NESSI). NESSI was funded by the NASA Exoplanet Exploration Program and the NASA Ames Research Center. NESSI was built at the Ames Research Center by Steve B. Howell, Nic Scott, Elliott P. Horch, and Emmett Quigley. The observations fall under the program number 2023B-438370 (PI: Kanodia). \\
\indent We extend our sincere appreciation to the staff of the RBO telescope facility whose expertise was vital in the success of this research paper.\\

\noindent This research has made use of the NASA Exoplanet Archive, which is operated by the California Institute of Technology, under contract with the National Aeronautics and Space Administration under the Exoplanet Exploration Program.\\
\indent This work has made use of data from the European Space Agency (ESA) mission
{\it Gaia}\footnote{\url{https://www.cosmos.esa.int/gaia}}, processed by the {\it Gaia} Data Processing and Analysis Consortium (DPAC\footnote{ \url{https://www.cosmos.esa.int/web/gaia/dpac/consortium}}). Funding for the DPAC has been provided by national institutions, in particular the institutions
participating in the {\it Gaia} Multilateral Agreement.\\
\indent This publication makes use of data products from the Two Micron All Sky Survey (2MASS), which is a joint project of the University of Massachusetts and the Infrared Processing and Analysis Center/California Institute of Technology, funded by the National Aeronautics and Space Administration and the National Science Foundation.\\
\indent This publication makes use of data products from the Wide-field Infrared Survey Explorer, which is a joint project of the University of California, Los Angeles, and the Jet Propulsion Laboratory/California Institute of Technology, and NEOWISE, which is a project of the Jet Propulsion Laboratory/California Institute of Technology. WISE and NEOWISE are funded by the National Aeronautics and Space Administration.\\
\indent This research was made possible through the use of the AAVSO Photometric All-Sky Survey (APASS), funded by the Robert Martin Ayers Sciences Fund and NSF AST-1412587.\\

\noindent {\it Facilities}: TESS, HET (HPF), RBO, Gaia, Nasa Exoplanet Archive, 2MASS, APASS, WISE, LRS-2, NESSI, ZTF, ASAS-SN.\\

\noindent All the TESS data used in this paper can be found in MAST: \dataset[10.17909/h9vj-t740]{http://dx.doi.org/10.17909/h9vj-t740}.


\bibliography{sample631}{}
\bibliographystyle{aasjournal}



\end{document}